\documentclass[traditabstract]{aa}
\usepackage{graphics}
\usepackage{graphicx}
\usepackage{multicol}
\usepackage[varg]{txfonts}

\usepackage{natbib,twoopt}
\usepackage{amsmath} 
\usepackage[breaklinks=true]{hyperref} 
\bibpunct{(}{)}{;}{a}{}{,} 
\makeatletter
\newcommandtwoopt{\citeads}[3][][]{\href{http://adsabs.harvard.edu/abs/#3}%
{\def\hyper@linkstart##1##2{}%
\let\hyper@linkend\@empty\citealp[#1][#2]{#3}}}
\newcommandtwoopt{\citepads}[3][][]{\href{http://adsabs.harvard.edu/abs/#3}%
{\def\hyper@linkstart##1##2{}%
\let\hyper@linkend\@empty\citep[#1][#2]{#3}}}
\newcommandtwoopt{\citetads}[3][][]{\href{http://adsabs.harvard.edu/abs/#3}%
{\def\hyper@linkstart##1##2{}%
\let\hyper@linkend\@empty\citet[#1][#2]{#3}}}
\newcommandtwoopt{\citeyearads}[3][][]%
{\href{http://adsabs.harvard.edu/abs/#3}
{\def\hyper@linkstart##1##2{}%
\let\hyper@linkend\@empty\citeyear[#1][#2]{#3}}}
\makeatother

\usepackage{color}
\definecolor{mygreen}{RGB}{0,128,0}

\hypersetup{colorlinks=true,linkcolor=blue,citecolor=blue,urlcolor=blue}

\def\thetaA{0.558} 
\def\thetaAerr{0.008 \pm 0.020} 
\def\thetaB{0.539} 
\def\thetaBerr{0.009 \pm 0.020} 
\def\uclaret{0.3361}
\def\kfactor{1.0262}
\def\thetaLDA{0.573} 
\def\thetaLDAerr{0.021} 
\def\thetaLDB{0.554} 
\def\thetaLDBerr{0.022} 
\def\radA{0.165} 
\def\radAerr{0.006} 
\def\radB{0.159} 
\def\radBerr{0.006} 
\def\Mtot{0.242}
\def\Mtoterr{0.006}
\def\MA{0.1225}
\def\MAerr{0.0043}
\def\MB{0.1195}
\def\MBerr{0.0043}
\def\loggA{$5.092 \pm 0.015$}
\def\loggB{$5.113 \pm 0.015$}

\begin{document}

\title{The red dwarf pair GJ65\,AB: inflated, spinning twins of Proxima}
\subtitle{Fundamental parameters from PIONIER, NACO, and UVES observations
\thanks{Based on observations collected at the European Organisation for Astronomical Research in the Southern Hemisphere under ESO programmes 076.B-0055(A), 173.C-0606(B), 079.C-0216(A), 082.C-0518(C), 086.C-0515(A), 087.D-0150(B) and 091.D-0074(A).}}
\titlerunning{Fundamental parameters of the binary red dwarf GJ65}
\authorrunning{P. Kervella et al.}
%
\author{
P.~Kervella\inst{1,2}
\and
A.~M\'erand\inst{3}
\and
C.~Ledoux\inst{3}
\and
B.-O.~Demory\inst{4}
\and
J.-B.~Le Bouquin\inst{5}
}
\institute{
Unidad Mixta Internacional Franco-Chilena de Astronom\'{i}a (CNRS UMI 3386), Departamento de Astronom\'{i}a, Universidad de Chile, Camino El Observatorio 1515, Las Condes, Santiago, Chile, \email{pkervell@das.uchile.cl}.
\and
LESIA (UMR 8109), Observatoire de Paris, PSL Research University, CNRS, UPMC, Univ. Paris-Diderot, 5 Place Jules Janssen, 92195 Meudon, France, \email{pierre.kervella@obspm.fr}.
\and
European Southern Observatory, Alonso de C\'ordova 3107, Casilla 19001, Santiago 19, Chile.
\and
Cavendish Laboratory, University of Cambridge, J. J. Thomson Avenue, Cambridge, CB30HE UK.
\and
UJF-Grenoble\,1/CNRS-INSU, Institut de Plan\'etologie et d'Astrophysique de Grenoble (UMR 5274), 38041 Grenoble, France.
}
\date{Received 3 April 2016; Accepted 20 June 2016.}
\abstract
{
The nearby red dwarf binary GJ65\,AB (UV+BL\,Ceti, M5.5Ve+M6Ve) is a cornerstone system to probe the physics of very low-mass stars. The radii of the two stars are currently known only from indirect photometric estimates, however, and this prevents us from using GJ65\,AB as calibrators for the mass-radius (M--R) relation.
We present new interferometric measurements of the angular diameters of the two components of GJ65 with the VLTI/PIONIER instrument in the near-infrared $H$ band: $\theta_\mathrm{UD}(\mathrm{A}) = \thetaA \pm \thetaAerr$\,mas and $\theta_\mathrm{UD}(\mathrm{B}) = \thetaB \pm \thetaBerr$\,mas. They translate into limb-darkened angular diameters of $\theta_\mathrm{LD}(\mathrm{A}) = \thetaLDA \pm \thetaLDAerr$\,mas and $\theta_\mathrm{LD}(\mathrm{B}) = \thetaLDB \pm \thetaLDBerr$\,mas.
Based on the known parallax, the linear radii are $R({\rm A}) = \radA \pm \radAerr\ R_\odot$ and $R({\rm B}) = \radB \pm \radBerr\ R_\odot$ ($\sigma(R)/R = 4\%$). We searched for the signature of flares and faint companions in the interferometric visibilities and closure phases, but we did not identify any significant signal.
We also observed GJ65 with the VLT/NACO adaptive optics and refined the orbital parameters and infrared magnitudes of the system. We derived masses for the two components of $m(\mathrm{A}) = \MA \pm \MAerr\,M_\odot$ and $m(\mathrm{B}) = \MB \pm \MBerr\,M_\odot$ ($\sigma(m)/m = 4\%$).
To derive the radial and rotational velocities of the two stars as well as their relative metallicity with respect to Proxima, we also present new individual UVES high-resolution spectra of the two components.
Placing GJ65\,A and B in the mass-radius diagram shows that their radii exceed expectations from recent models by $14 \pm 4\%$  and $12 \pm 4\%$ , respectively.
Following previous theories, we propose that this discrepancy is caused by the inhibition of  convective energy transport by a strong internal magnetic field generated by dynamo effect in these two fast-rotating stars. A comparison with the almost identical twin Proxima, which is rotating slowly, strengthens this hypothesis because the radius of Proxima does not appear to be inflated compared to models.}

\maketitle

\newpage

\section{Introduction}

Stellar structure models have traditionally been found to systematically underestimate the radii of very low-mass stars (VLMS) of a given mass by 5 to 15\% \citepads{2002ApJ...567.1140T, 2010ApJ...718..502M}. 
To explain the discrepancy between models and observations, \citetads{2007A&A...472L..17C} proposed that a decrease in convection efficiency induced by strong magnetic fields can inflate the stellar radius by amounts that qualitatively match the observed differences.
Recent progress has been reported by \citetads{2014ApJ...789...53F} using the Dartmouth models \citepads{2008ApJS..178...89D}. 
The differences between models and eclipsing binary observations are now lower than in the past,~within a few percent for \object{CM Dra} \citepads{2014A&A...571A..70F}, for instance, and the average discrepancy is around 3\% \citepads{2013ApJ...776...87S}. However, there are still only very few test stars for these models around $0.1\,M_\odot$, at the limit of the brown dwarf regime.
A thorough review of the field of low-mass star evolution modeling can be found in \citetads{2015ASPC..496..137F}.

To validate the VLMS structure models, we thus need radius measurements at the low-mass end of the stellar mass-radius (M--R) function.
As shown by \citetads{2003A&A...397L...5S}, \citetads{2009A&A...505..205D}, and \citetads{2012ApJ...757..112B}, for example, long-baseline interferometry is well suited for this task because it provides a simple way to measure the angular diameter of non-eclipsing stars.
Unfortunately, the number of VLMS accessible to interferometric measurements is limited by their very small physical radius.
With the current optical and infrared interferometry instrumentation, the largest distance at which red dwarfs can be resolved angularly is only about 3\,pc.
For this reason, eclipsing VLMS binaries are also used as calibrators for the M--R relation, but as a result of the proximity of the stars, the gravitational and magnetic interactions can affect their physical properties. \citetads{2013ApJ...776...87S} found a comparable discrepancy level on average of observed vs.~model radii for single and binary VLMS, but the components of short-period binaries were found to be the most deviant. Observations of single stars and well-separated binaries is therefore very desirable to calibrate the M--R relation.

Its wide physical separation (larger than 4\,AU for most of the orbit) and proximity ($d = 2.7$\,pc) make the \object{Gliese 65} system (\object{GJ65} AB, \object{Luyten 726-8}, \object{BL Cet}+\object{UV Cet}, \object{WDS J01388-1758AB}, \object{2MASS J01390120-1757026}, \object{LDS 838}) a cornerstone system on
which to calibrate the M--R relation.
This binary consists of two main-sequence red dwarfs of spectral types M5.5Ve and M6Ve \citepads{1994AJ....108.1437H}. Its relatively wide separation is sufficient to neglect gravitational and magnetic interactions, but the two stars are close enough so that the period is relatively short ($P=26.3$\,years). Accurate orbital parameters and masses can therefore be determined in a reasonable time frame.
These ideal properties make GJ65\,AB a rare and very favorable configuration among the known VLMS systems. As a side remark, GJ65\,AB will pass within one\,light-year of $\epsilon$\,Eri in about 30\,000\,years \citepads{2010arXiv1004.1557P}, possibly interacting with the hypothetical Oort cloud of this star.

We present our new observations in Sect.~\ref{observations}. We used the VLTI/PIONIER instrument to measure the angular diameters of the two components of GJ65 (Sect.~\ref{pionier_obs}), and we combined them with the well-known parallax of the system to derive their linear radius.
We also observed GJ65 using NACO (Sect.~\ref{naco_obs}), from which we confirm that a revision of the orbital parameters is necessary.
In Sect.~\ref{uves_data} we present high-resolution spectra of the two stars.
The determination of the parameters of GJ65 A and B from our observations is described in Sect.~\ref{analysis}.
We derive an improved orbital solution in Sect.~\ref{orbital-solution}, including a revised estimate of the mass of the system.
We finally compare GJ65\,AB with \object{Proxima} in Sect.~\ref{proxima}. The fundamental parameters of these three stars are very similar, and we discuss the possible reasons for the discrepancy in their measured radii.

\section{Observations and data reduction}\label{observations}

\subsection{VLTI/PIONIER interferometry \label{pionier_obs}}

\begin{figure*}[]
        \centering
        \includegraphics[width=9cm]{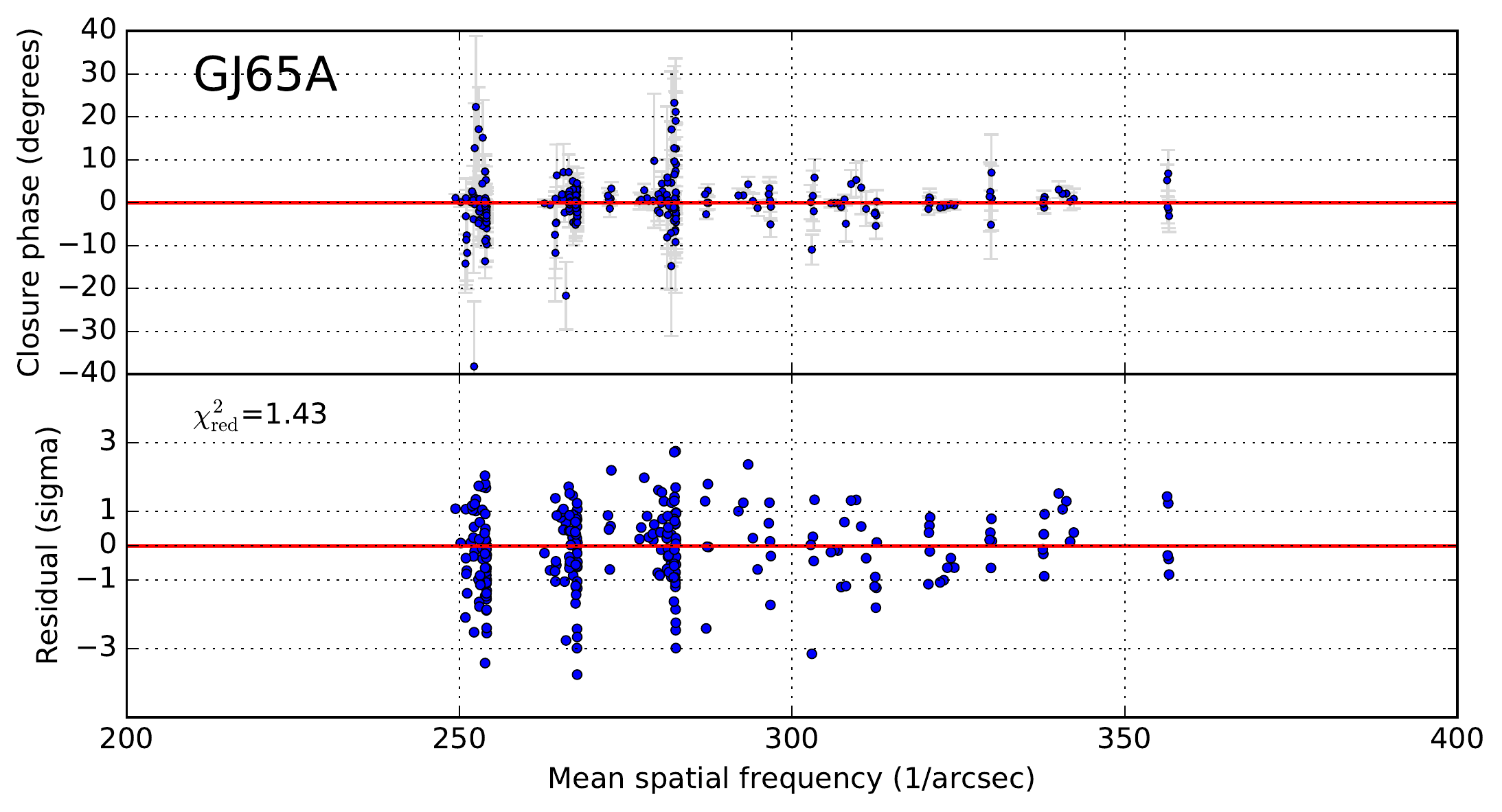}
        \includegraphics[width=9cm]{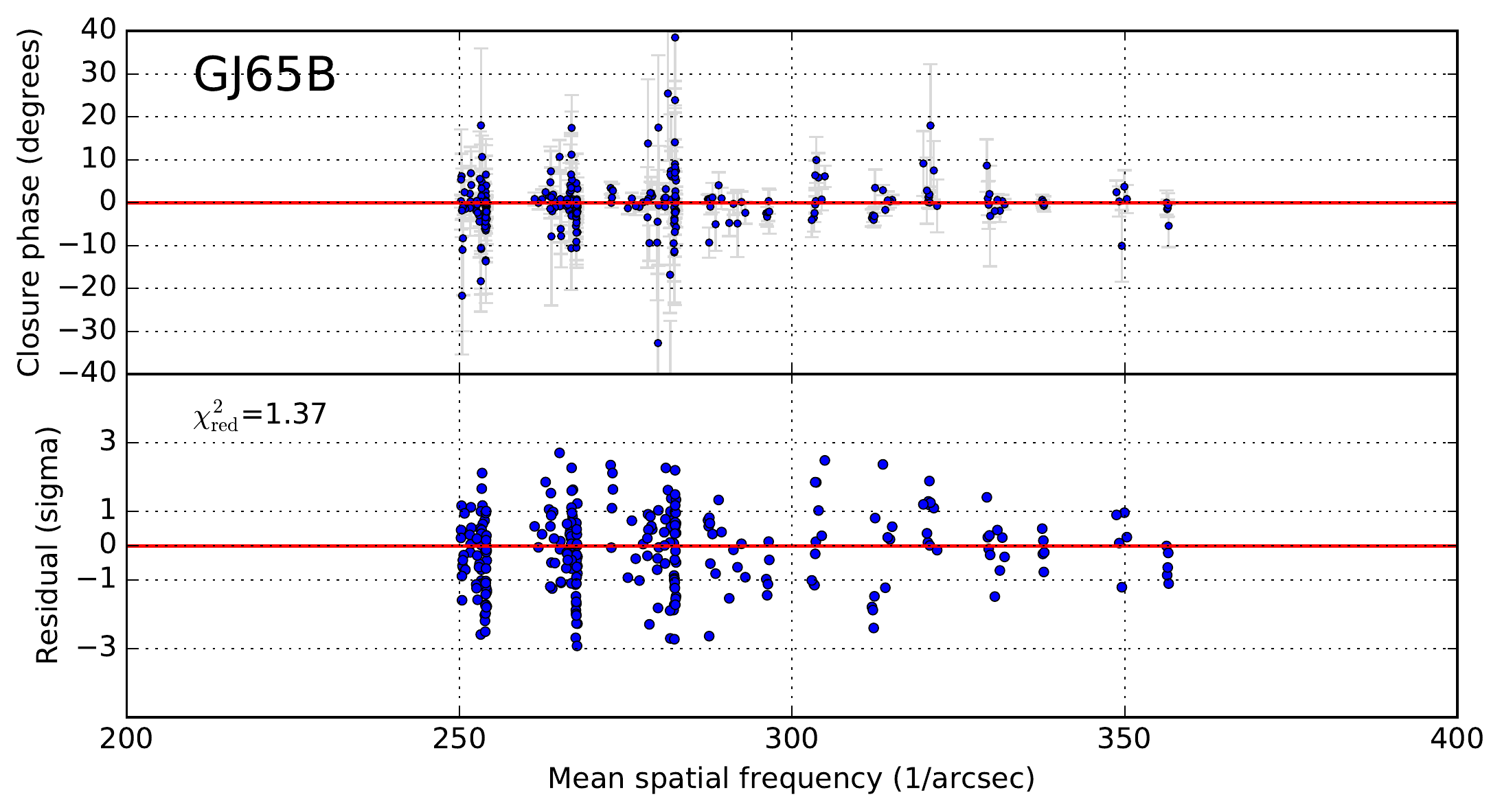}
        \includegraphics[width=9cm]{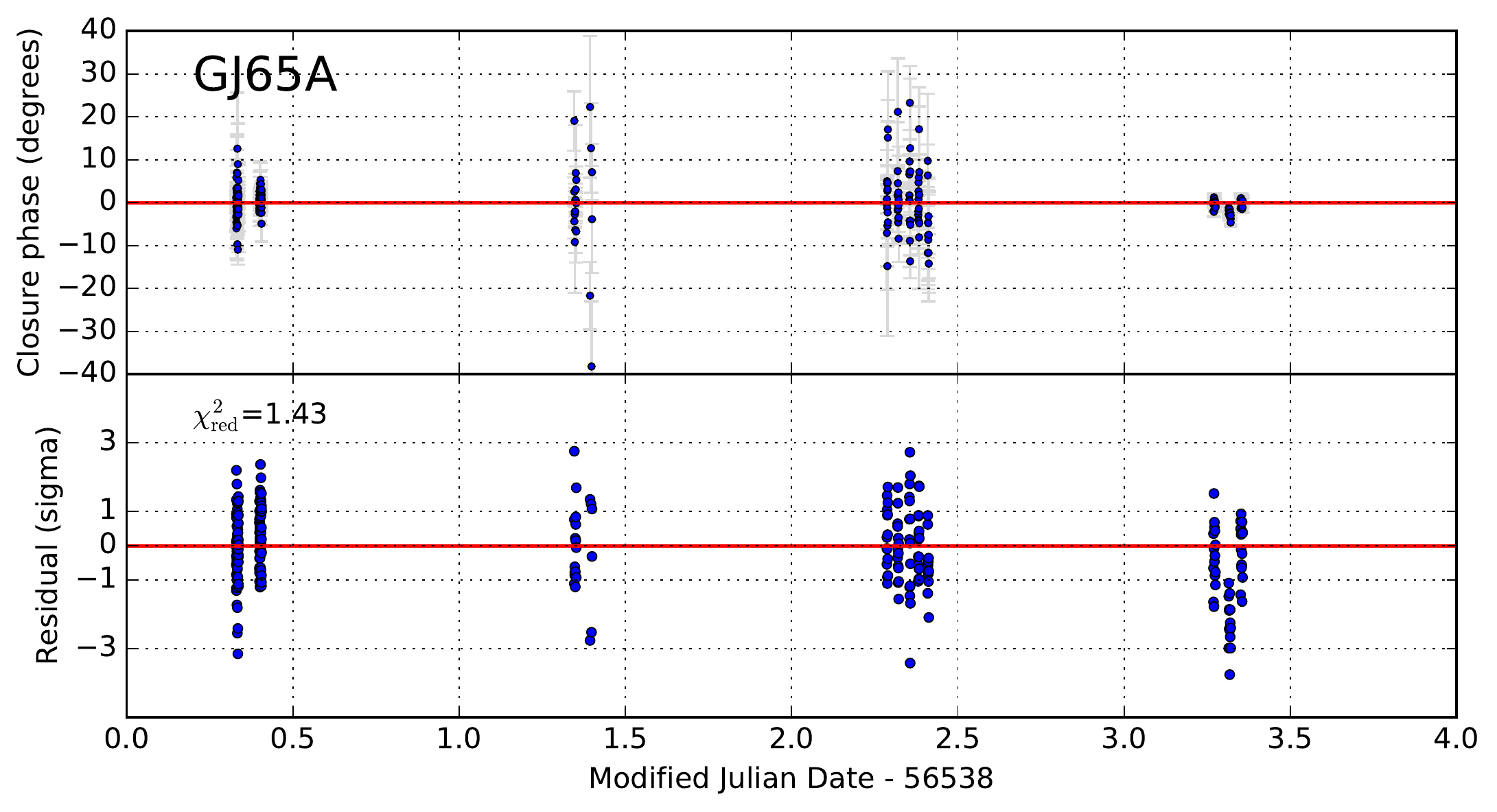}
        \includegraphics[width=9cm]{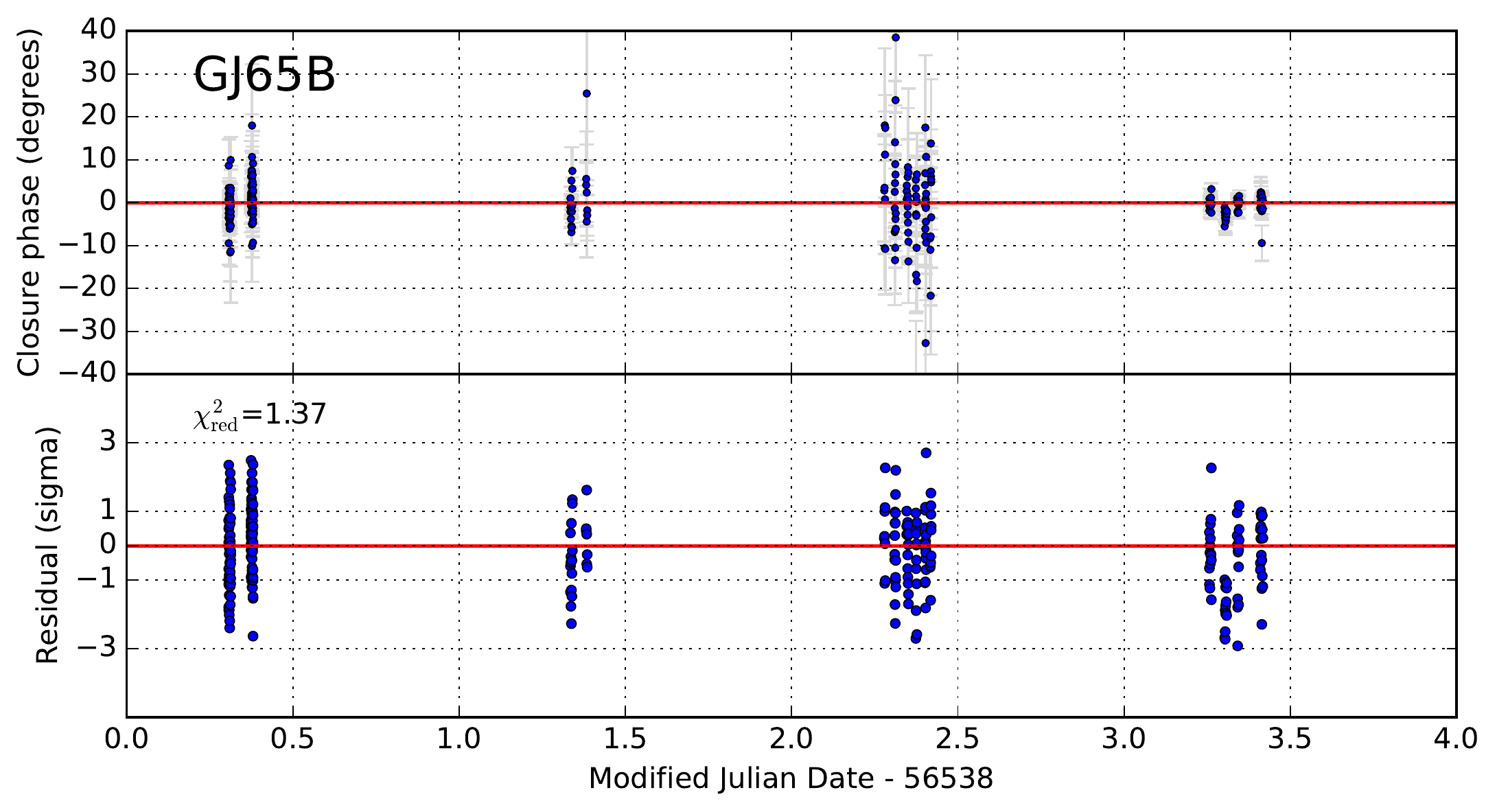}
        \caption{Closure phases of GJ65\,A and B as a  function of the mean spatial frequency of the baseline triangle (top row) and of time (bottom row). The residuals with respect to zero are shown in the bottom panels. \label{fig-t3phi}}
\end{figure*}

We observed GJ65\,A and B during four consecutive nights between 2 September and 5 September 2013 (UT dates), using the Very Large Telescope Interferometer \citepads{2014SPIE.9146E..0JM} equipped with the four-telescope beam combiner PIONIER (\citeads{2010SPIE.7734E..99B}, \citeads{2011A&A...535A..67L}). We used the A1-G1-K0-J3 quadruplet, which provides ground-baseline lengths between 57 and 140\,meters.
These baselines resolve GJ65\,A and B only moderately, but longer baselines are currently not available at VLTI. We selected the $H$ -band filter of PIONIER and adopted the low spectral resolution (three wavelength channels at 1.590, 1.678, and 1.768\,$\mu$m) and the Fowler detector readout mode.
The adopted interferometric calibrator stars are listed in Table~\ref{calibrators}. The seeing in the visible varied from 0.6 to 1.6\,arcseconds during the observing run.
Owing to a technical problem with one of the delay lines on 4 September 2013, the observations were obtained with only three telescopes, resulting in a single closure phase measurement. 
The interferometric closure phases measured with PIONIER are shown in Fig.~\ref{fig-t3phi}. They do not show a significant residual above the noise level, either as a function of spatial frequency or as a function of time.
 
\begin{figure*}[]
        \centering
        \includegraphics[width=9cm]{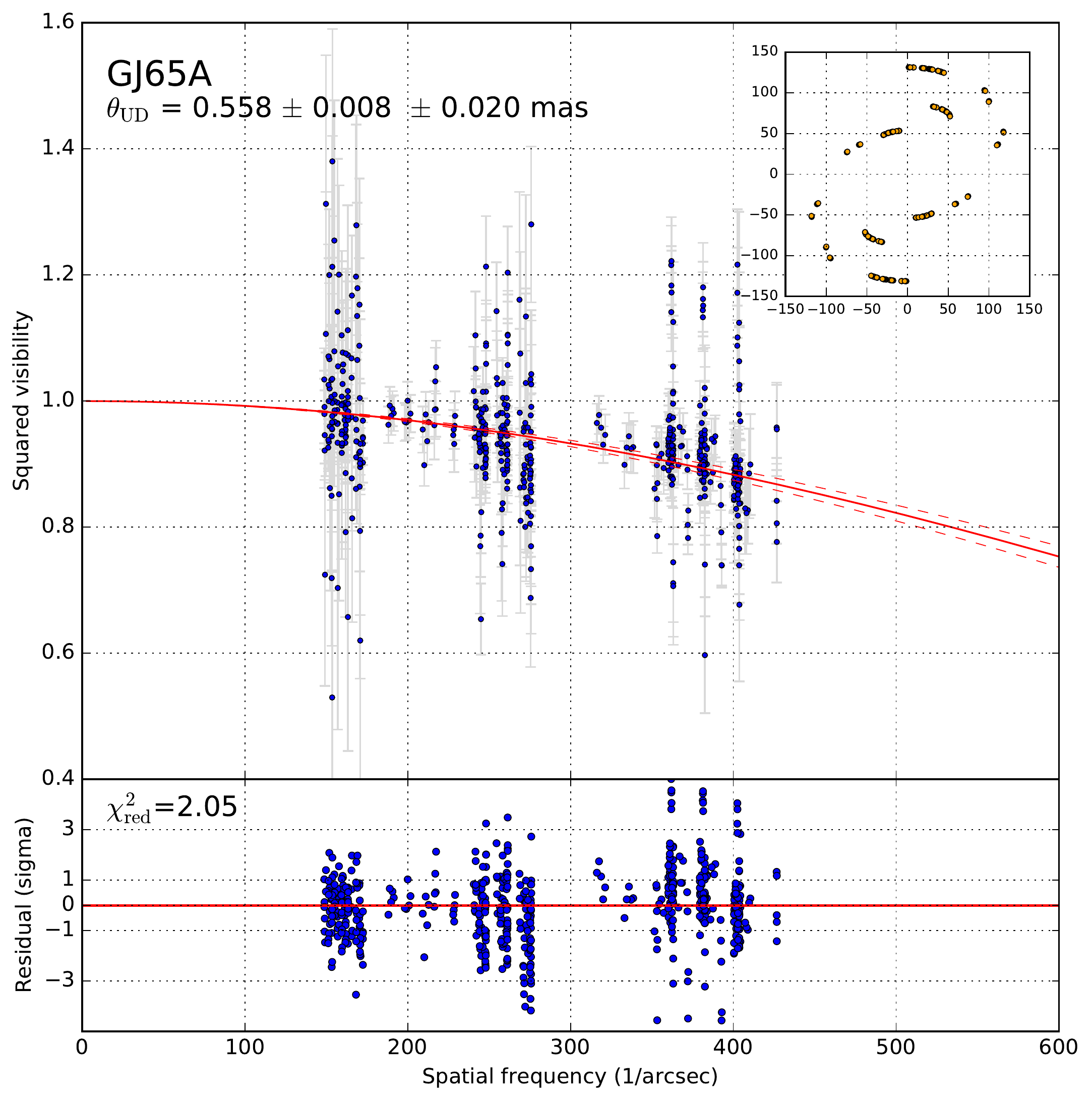}
        \includegraphics[width=9cm]{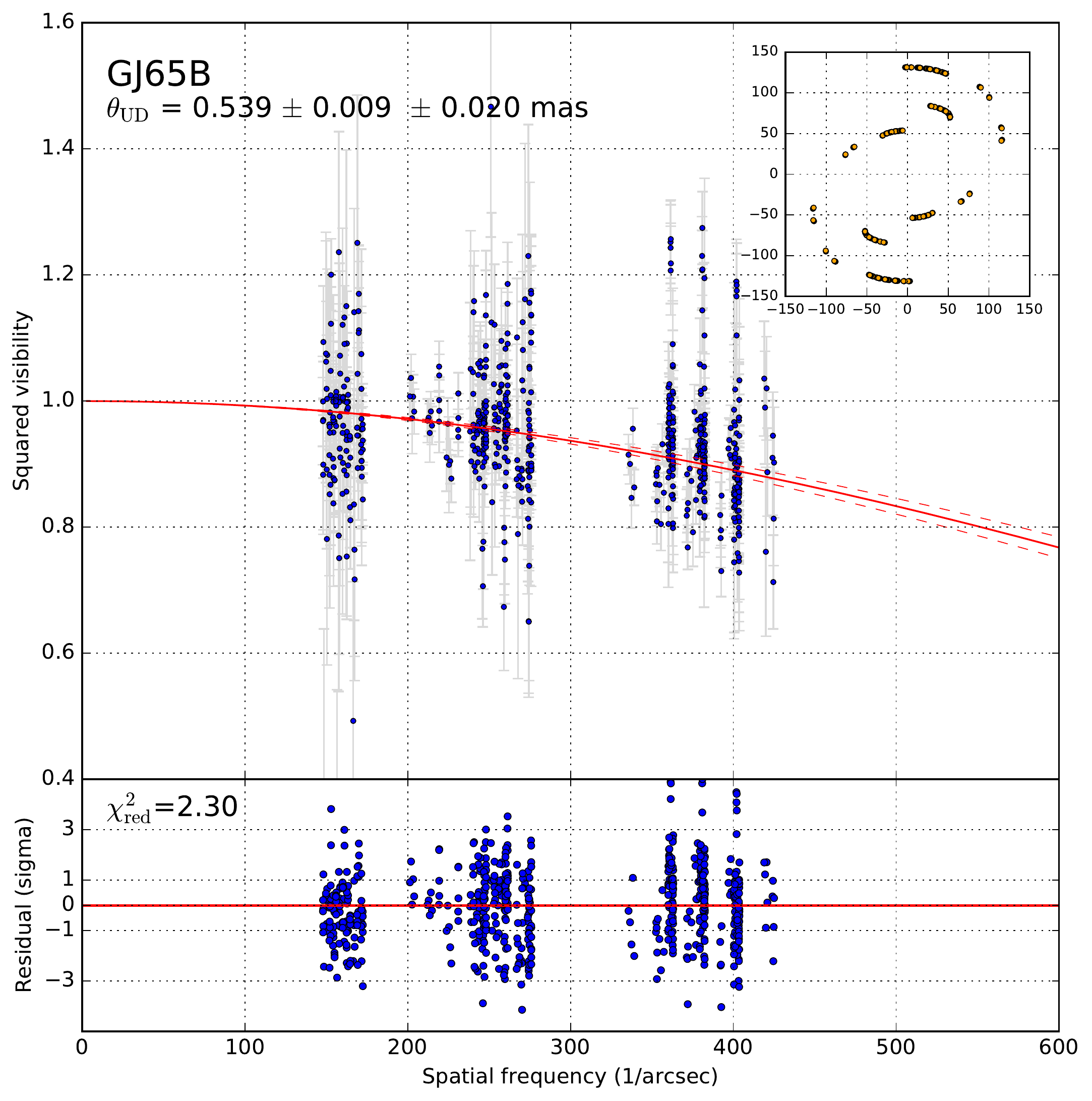}
        \caption{Squared visibilities of GJ65\,A and B and adjusted uniform disk models. The statistical dispersion and the calibration systematic uncertainty are listed separately after the best-fit angular diameter. The dashed lines delimit the 1$\sigma$ total uncertainty of the fit. The $(u,v)$ plane coverage is shown in the top right panel, with axes labeled in meters. The residuals of the model fit are shown in the bottom panels, and the reduced $\chi^2$ values are computed considering only the statistical uncertainties.\label{fig-visibilities}}
\end{figure*}

The measured visibilities and adjusted uniform disk (UD) visibility models for the two stars are presented in Fig.\,\ref{fig-visibilities}.
We derive UD angular diameters of $\theta_\mathrm{UD}(\mathrm{A}) = \thetaA \pm \thetaAerr$\,mas and $\theta_\mathrm{UD}(\mathrm{B}) = \thetaB \pm \thetaBerr$\,mas. We list the statistical dispersion and the systematic calibration uncertainty separately. The reduced $\chi^2$ of the uniform disk angular diameter fit is around 2 for the two stars considering the statistical dispersion alone. Considering the relatively large systematic calibration uncertainty compared to the statistical error bar, we did not renormalize the statistical error bars.
We transformed the UD angular diameters into limb-darkened (LD, i.e.,\ photospheric) angular diameters using the linear limb-darkening coefficient derived from quasi-spherical PHOENIX model atmospheres by \citetads{2012A&A...546A..14C} for $T_\mathrm{eff} = 3000$\,K, $v_\mathrm{turb}=2.0$\,km.s$^{-1}$, $\log g = 4.5$ \citepads{2005MNRAS.358..105J} giving $u_K=\uclaret$ (flux conservation method). We then computed the ratio $k = \theta_{\rm {LD}} / \theta_{\rm {UD}} = \kfactor$ following the prescription by \citetads{1974MNRAS.167..475H}.
This gives LD angular diameters of $\theta_\mathrm{LD}(\mathrm{A}) = \thetaLDA \pm \thetaLDAerr$\,mas and $\theta_\mathrm{LD}(\mathrm{B}) = \thetaLDB \pm \thetaLDBerr$\,mas.
Considering the parallax $\pi=373.70 \pm 2.70$\,mas ($\pm 0.7\%$ accuracy) from \citetads{1995gcts.book.....V} (see also \citeads{1995yCat.1174....0V}), this corresponds to linear photospheric radii of $R({\rm A}) = \radA \pm \radAerr\ R_\odot$ and $R({\rm B}) = \radB \pm \radBerr\ R_\odot$.
We note that the parallax from \citetads{1995gcts.book.....V} is compatible with the value of $\pi=375 \pm 4$\,mas published by \citetads{1988AJ.....95.1841G}
From our radius and mass estimates, we derive mean densities of $\rho_A = 3.8\ 10^{4}$\,kg.m$^{-3}$ and $\rho_B = 4.2\ 10^{4}$\,kg.m$^{-3}$. As a basis for comparison, these remarkably high values correspond to approximately 30 times the mean density of the Sun, or twice the density of metallic gold.

GJ65 A and B are both relatively fast-rotating stars, as we derive projected rotational velocities $v \sin i$ around 30\,km\,s$^{-1}$ (Sect.~\ref{rotvel}).
According to the catalog assembled by \citetads{2009ApJ...704..975J}, this fast rotational velocity is well above the typical values for M5-M6 dwarfs ($\approx 8$\,km/s). The activity saturation limits determined by these authors indicate that the system is relatively young, probably a few billion years at most. We might speculate, however, that the magnetic wind-braking mechanism is less efficient in a binary system with two magnetic stars than for single stars because of the perturbation of the wind flow by the second star. This would result in a faster rotation velocity for a star in a binary system than for a single star of the same age.
The fast rotation of GJ65\,AB, however, does not induce any noticeable flattening of their photosphere, as a simple calculation using the Huygens approximation gives a negligible expected flattening ratio $f$ (assuming $v \sin i = 30$\,km/s):
\begin{equation}
f = \frac{R_\mathrm{eq}-R_\mathrm{pol}}{R_{pol}} \approx \frac{3}{8 \pi G} \frac{\omega^2}{\rho} = 0.3\%.
\end{equation}

\begin{table*}
        \caption{Calibrators for the PIONIER observations of GJ65\,AB. `Dist.' is the angular distance of each calibrator to GJ65\,AB.}
        \centering          
        \label{calibrators}
        \begin{tabular}{lccclccll}
        \hline\hline
        \noalign{\smallskip}
        HD & Dist. & $m_V$ & $m_H$ & Spect. & $\theta_\mathrm{LD}$ & $\theta_\mathrm{UD}(H)$ & Ref. & Dates \\
         & $[\deg]$ &  & & & [mas] & [mas]  & & [Sept. 2013] \\
        \noalign{\smallskip}
        \hline    
        \noalign{\smallskip}
        6482 & 11.4 & 6.10 & 3.81 & K0III & $0.859 \pm 0.012$ & $0.836 \pm 0.012$ & M05 & 03, 04 \\
        8959 & 4.8 & 6.56 & 4.66 & K0III & $0.607 \pm 0.042$ & $0.590 \pm 0.042$& SearchCal & 04, 05 \\
        10148 & 3.3 & 5.57 & 4.83 & F0V & $0.444 \pm 0.031$ & $0.437 \pm 0.031$& SearchCal & 04, 05 \\
        13004 & 6.7 & 6.39 & 3.89 & K1III & $0.850 \pm 0.059$ & $0.824 \pm 0.059$& SearchCal & 02, 03, 04  \\
        \hline                      
        \end{tabular}
        \tablefoot{M05: \citetads{2005A&A...433.1155M}, SearchCal: \citetads{2006A&A...456..789B, 2011A&A...535A..53B}.
        }
\end{table*}

\subsection{VLT/NACO astrometry and photometry \label{naco_obs}}

\begin{figure}
        \centering
        \includegraphics[width=4.4cm,page=1]{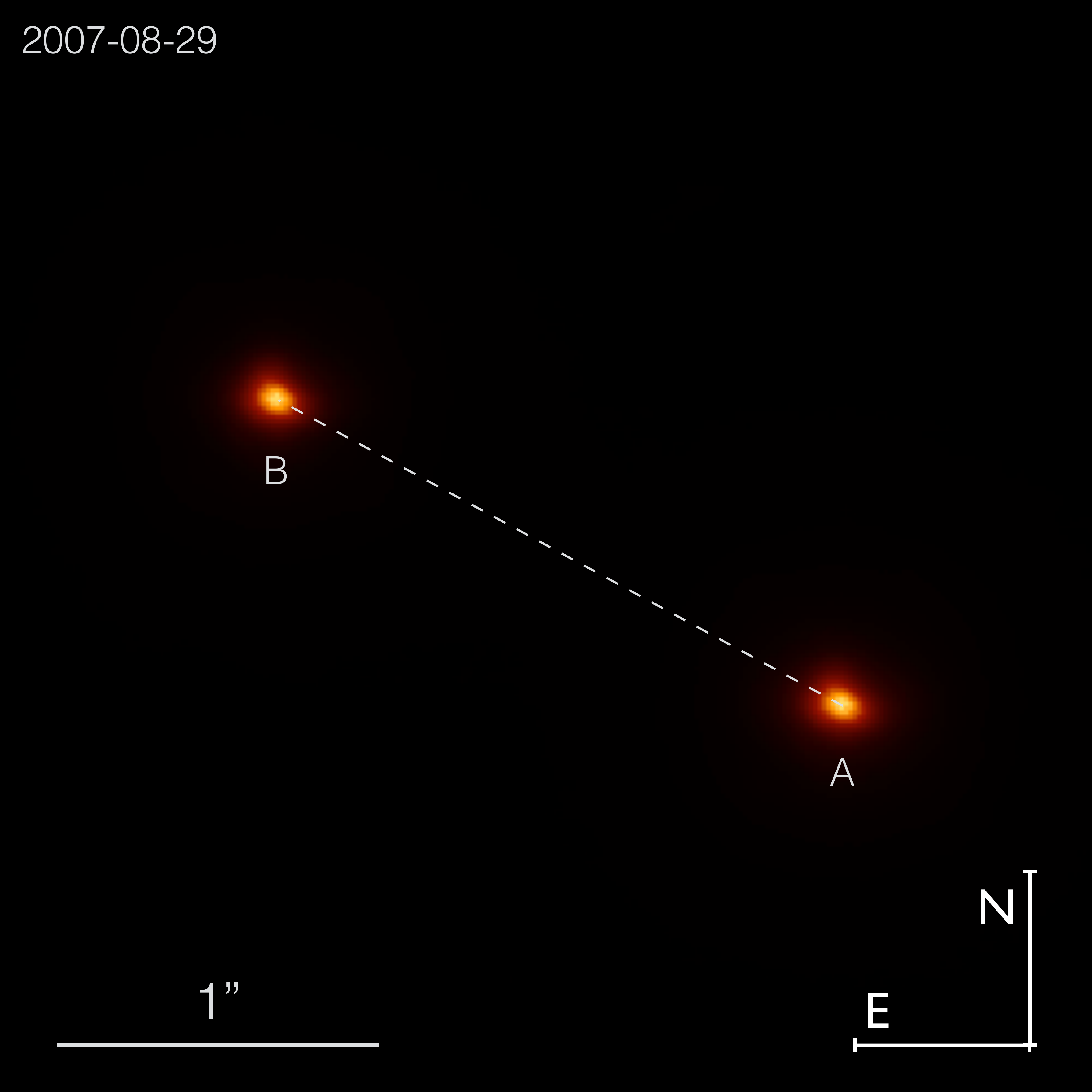}
        \includegraphics[width=4.4cm,page=2]{Figures/FigSquared.pdf}
        \caption{NACO images of GJ65 at two epochs. \label{naco-epochs}}
\end{figure}

The orbital elements and masses of GJ65 were published by \citetads{1973AJ.....78..650W} and \citetads{1988AJ.....95.1841G}.
However, high-accuracy astrometry obtained by \citetads{2000AJ....119..906S} using the \emph{Hubble Space Telescope} Wide Field and Planetary Camera 2 (HST/WFPC2) showed a significant discrepancy with the published orbit, and the masses were tentatively revised to 0.115 and 0.113\,M$_\odot$.
Our observation of GJ65 on 23 September 2011 with the VLT/NACO adaptive optics system \citepads{2003SPIE.4839..140R} confirmed that a revision of the orbital parameters is necessary.
We therefore retrieved the available NACO observations of GJ65 from the ESO archive; they were obtained between 2002 and 2011. They cover five epochs in total, including our own 2011 observation. As the orbital period is around 26\,years, the relative displacement of the two stars is easily measurable from epoch to epoch (Fig.~\ref{naco-epochs}). The observations of GJ65 were obtained with the short-wavelength S13 and the long-wavelength L27 cameras. We processed the raw data and measured the relative  position of GJ65\,B with respect to GJ65\,A. The NACO astrometric calibration for the S13 camera is taken from \citetads{2008A&A...484..281N}: the plate scale is $13.26 \pm 0.06$\,mas/pix, and the position angle of the Y axis of the detector from North to East is $+0.34 \pm 0.38$\,degrees. The astrometric calibration for the L27 camera is taken from \citetads{2006A&A...456.1165C}: the plate scale is $27.01 \pm 0.05 $\,mas/pix, and the position angle is $+0.00 \pm 0.20$\,degrees. These values are stable over several years within their stated error bars, and they agree with the calibrations of \citetads{2003A&A...411..157M} and \citetads{2015A&A...573A.127C}.
The resulting differential astrometric measurements of the position of GJ65\,B relative to A are presented in Table~\ref{table-astrom} together with the high-precision HST astrometry by \citetads{2000AJ....119..906S} and \citetads{1538-3881-144-2-64}.

The image of GJ65\,B obtained by \citetads{1998A&A...331..596B} using very long-baseline radio interferometry showed that radio flares can extend up to a distance of 2\,mas, which is~several times the photospheric size of the star.
But as the photometric amplitude of the flares in the infrared is very small, we expect a negligible effect on the NACO astrometry. It will probably be much smaller than the star's angular diameter, which is negligible compared to our measurement errors. At visible wavelengths, the photometric amplitude of the flares is larger, but even considering a particularly strong flare, it is unlikely that the center of light of the star is displaced by more than $\approx 1$\,mas, which means that it is~well within the uncertainties of the WFPC2 astrometry (Table~\ref{table-astrom}).

\begin{table*}
        \caption{Astrometric position of GJ65\,B relative to A. For the new NACO measurements that we report, the statistical and systematic uncertainties are listed separately (detector plate scale and orientation on sky).}
        \centering          
        \label{table-astrom}
        \begin{tabular}{llccll}
        \hline\hline
        \noalign{\smallskip}
        Date \& UT & MJD & $\rho$ $[\arcsec]$ & $\theta$ $[\deg]$ (E of N) & Ref. & Instrument \\
        \noalign{\smallskip}
        \hline    
        \noalign{\smallskip}
        1997-06-25 UT23:00 & 50624.9583 & $0.796 \pm 0.007$ & $294.68 \pm 0.85$ & S00 & HST/WFPC2 \\
        1997-07-27 UT21:20 & 50656.8889 & $0.749 \pm 0.007$ & $290.62 \pm 0.83$ & S00 & HST/WFPC2 \\
        2002-11-08 UT22:17 & 52586.9285 & $1.653 \pm 0.008$ & $103.30 \pm 0.20$ & D12 & HST/NICMOS \\
        2007-08-29 UT08:15 & 54341.3445 & $1.930 \pm 0.003 \pm 0.009$ & $61.24 \pm 0.02 \pm 0.38$ & K16 & NACO S13 \\
        2008-10-17 UT05:07 & 54756.2139 & $1.975 \pm 0.005 \pm 0.004$ & $53.11 \pm 0.03 \pm 0.20$ & K16 & NACO L27 \\
        2008-11-16 UT00:56 & 54786.0393 & $1.987 \pm 0.006 \pm 0.004$ & $52.62 \pm 0.07 \pm 0.20$ & K16 & NACO L27 \\
        2010-10-16 UT04:33 & 55485.1899 & $2.095 \pm 0.006 \pm 0.011$ & $39.36 \pm 0.02 \pm 0.38$ & K16 & NACO S13 \\
        2011-09-23 UT06:28 & 55827.2696 & $2.134 \pm 0.001 \pm 0.010$ & $33.47 \pm 0.01 \pm 0.38$ & K16 & NACO S13 \\
        \hline                      
        \end{tabular}
        \tablefoot{
        K16: present work;
        S00: \citetads{2000AJ....119..906S};
        D12: \citetads{1538-3881-144-2-64}.
        }
\end{table*}

We list in Table~\ref{table-magdiff} the difference in magnitude of GJ65 A and B and the individual magnitudes measured from the NACO images in the infrared $JHK_s L'$ bands. We adopt the NACO photometric zero points from the instrument quality control database, which is available through the ESO archive web site\footnote{\url{http://www.eso.org/observing/dfo/quality/NACO/}}. The combined magnitudes we derive for A and B in the $JHK$ bands ($m_J=6.42$, $m_H=5.84$, $m_K=5.52$) are systematically fainter by $\approx 0.15$\,mag than the 2MASS values  ($m_J=6.28 \pm 0.02$, $m_H=5.69 \pm 0.03$, $m_K=5.34 \pm 0.02$; \citeads{2006AJ....131.1163S}). This might be due to a long-term variability of the two stars. The 2MASS observations were collected from Cerro Tololo in August 1998, as the two stars were significantly closer to each other ($\approx 1\arcsec$) than during our observations ($\approx 2\arcsec$). As a result of this proximity, the two stars were possibly more active \citepads{2011Ap.....54..469M} and brighter in the infrared at the time of the 2MASS observations.

Based on the combination of the $V$ magnitudes listed by \citetads{2001AJ....121.2189O}, $V(\mathrm{A})=12.52$ and  $V(\mathrm{B})=12.56$) and our NACO $K$ magnitudes, we can apply the surface brightness-color relations calibrated by \citetads{2004A&A...426..297K} to predict the angular diameters of the two stars. We obtain predicted angular diameters of $\theta_\mathrm{LD}(\mathrm{A})= 0.57 \pm 0.02$\,mas and $\theta_\mathrm{LD}(\mathrm{B})= 0.52 \pm 0.02$\,mas. These predicted values from photometry agree well with the PIONIER measurements ($\theta_\mathrm{LD}(\mathrm{A}) = \thetaLDA \pm \thetaLDAerr$\,mas and $\theta_\mathrm{LD}(\mathrm{B}) = \thetaLDB \pm \thetaLDBerr$\,mas; Sect.~\ref{pionier_obs}), strengthening our confidence in both the measured NACO magnitudes and the PIONIER angular diameters.

\begin{table*}
        \caption{NACO magnitudes of GJ65 A and B. The absolute magnitudes are computed using $\pi=373.70  \pm 2.70$\,mas \citepads{1995gcts.book.....V}.}
        \centering          
        \label{table-magdiff}
        \begin{tabular}{llccccccc}
        \hline\hline
        \noalign{\smallskip}
        UT Date & Filter & $\lambda_0$ [$\mu$m] & $m_B-m_A$ & NACO ZP & $m_\lambda(A)$ & $m_\lambda(B)$ & $M_\lambda(A)$ & $M_\lambda(B)$ \\
        \noalign{\smallskip}
        \hline    
        \noalign{\smallskip}
        2010-10-16 & $J$ & 1.27 & $0.156 \pm 0.004$ & $24.30 \pm 0.05$ & $7.10 \pm 0.05$ & $7.26 \pm 0.05$ & $9.96 \pm 0.05$ & $10.12 \pm 0.05$ \\
        2007-08-29 & NB$1.64$ & 1.64 & $0.148 \pm 0.001$ & -- & -- & -- & -- & -- \\
        2010-10-16 & $H$ & 1.66 & $0.150 \pm 0.007$ & $24.07 \pm 0.05$ & $6.50 \pm 0.05$ & $6.65 \pm 0.05$ & $9.36 \pm 0.05$ & $9.51 \pm 0.05$ \\
        2011-09-23 & $H$ & 1.66 & $0.171 \pm  0.006$ & $23.95 \pm 0.05$ &  $6.53 \pm 0.05$ & $6.70 \pm 0.05$ & $9.39 \pm 0.05$ & $9.56 \pm 0.05$  \\
        2011-09-23 & $K_s$ & 2.18 & $0.162 \pm  0.006$ & $23.09 \pm 0.05$ & $6.20 \pm 0.05$ & $6.36 \pm 0.05$ & $9.06 \pm 0.05$ & $9.22 \pm 0.05$  \\
        2008-10-17 & $L'$ & 3.80 & $0.102 \pm 0.002$ & $22.10 \pm 0.07$ & $5.92 \pm 0.07$ & $6.02 \pm 0.07$ & $8.78 \pm 0.07$ & $8.88 \pm 0.07$ \\
        2008-11-16 & $L'$ & 3.80 & $0.096 \pm 0.007$ & $22.10 \pm 0.07$ & $5.85 \pm 0.08$ & $5.94 \pm 0.08$ & $8.71 \pm 0.08$ & $8.80 \pm 0.08$ \\
        \hline
        \end{tabular}
        \tablefoot{
        $\lambda_0$ is the central wavelength of the filter, and ZP is the adopted NACO photometric zero point.
        }
\end{table*}

\subsection{VLT/UVES high-resolution spectroscopy\label{uves_data}}

\begin{figure}
        \centering
        \includegraphics[width=\hsize]{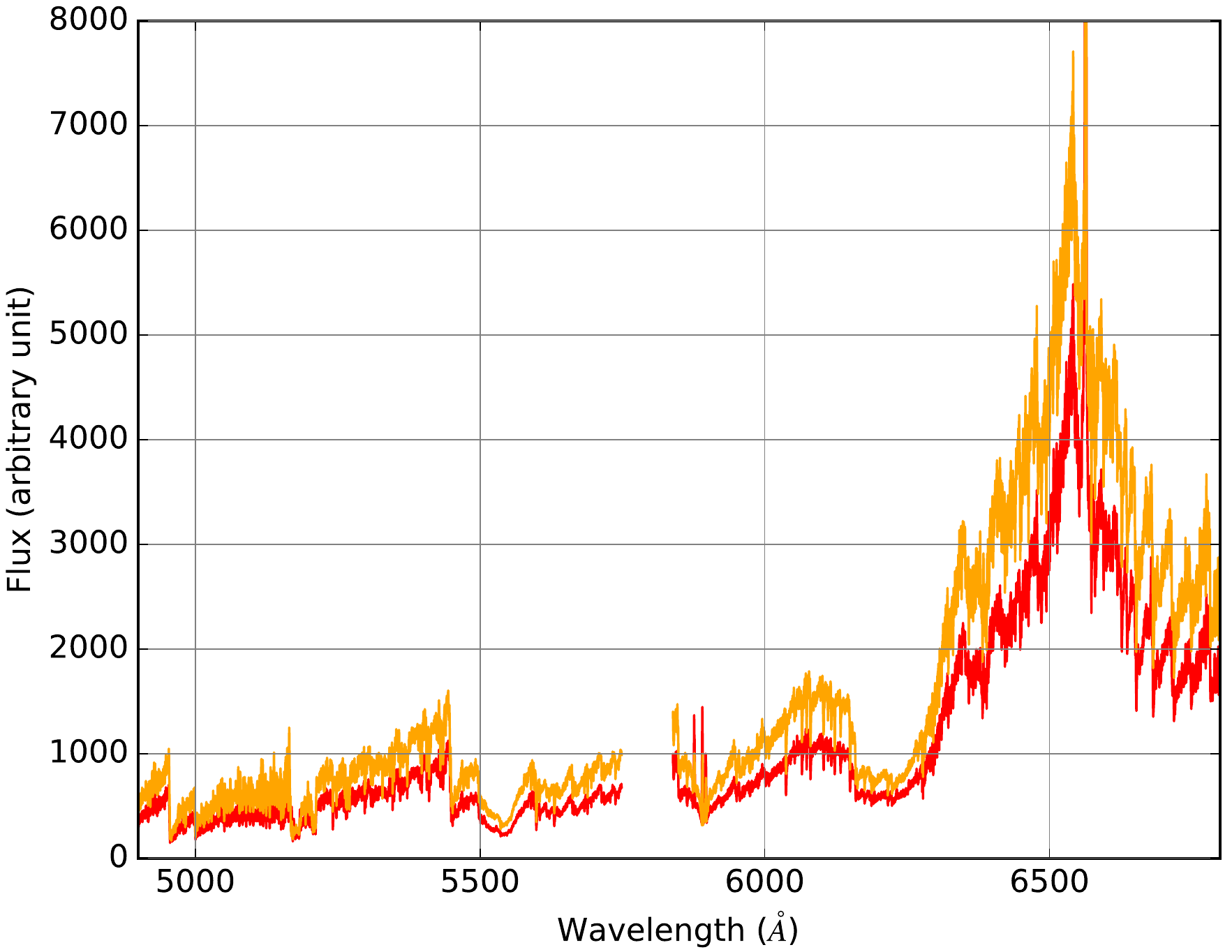}
        \caption{Overview of the UVES spectra of GJ65 A (orange) and B (red). \label{uves_global}}
\end{figure}

\begin{figure}
        \centering
        \includegraphics[width=8cm]{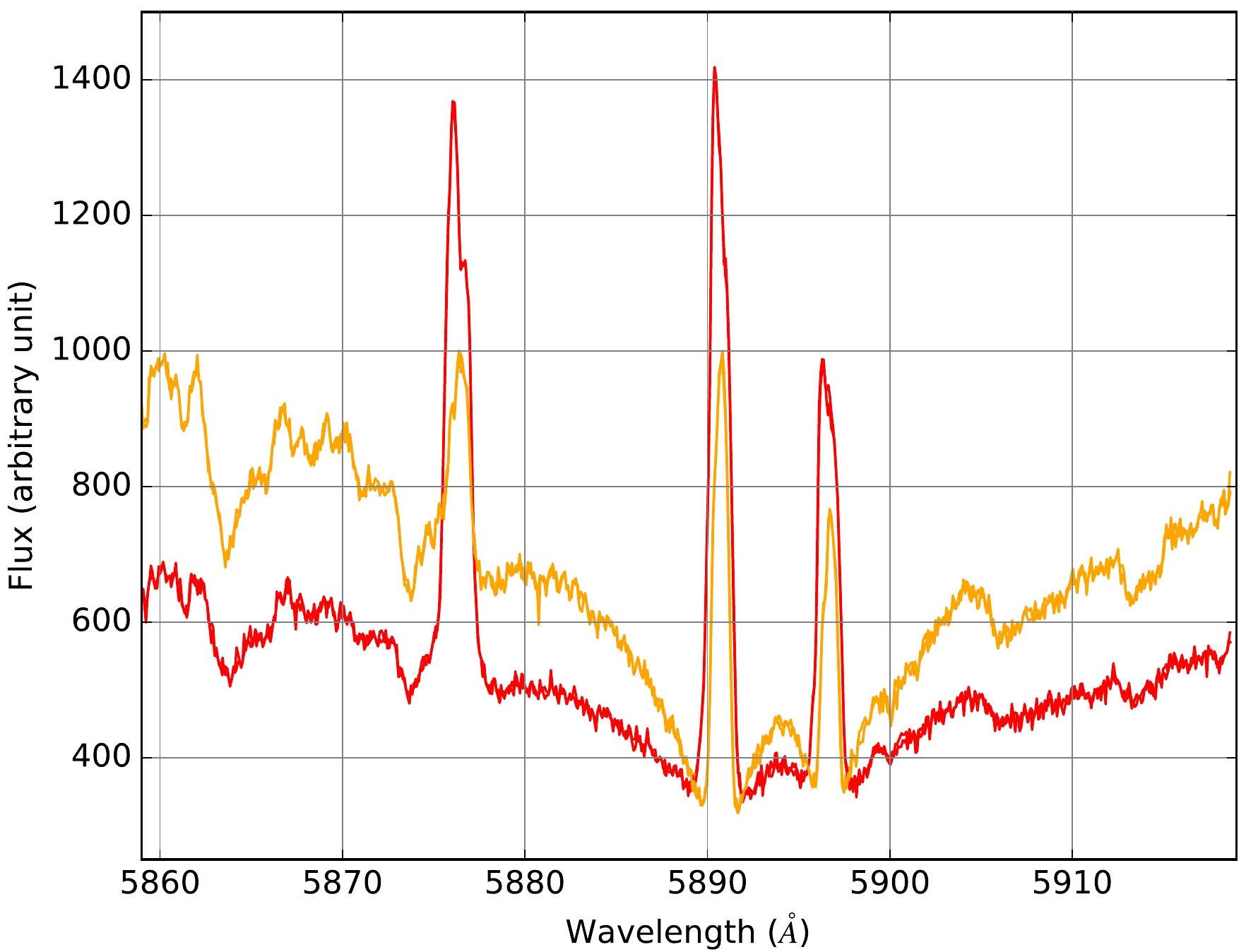}
        \includegraphics[width=8cm]{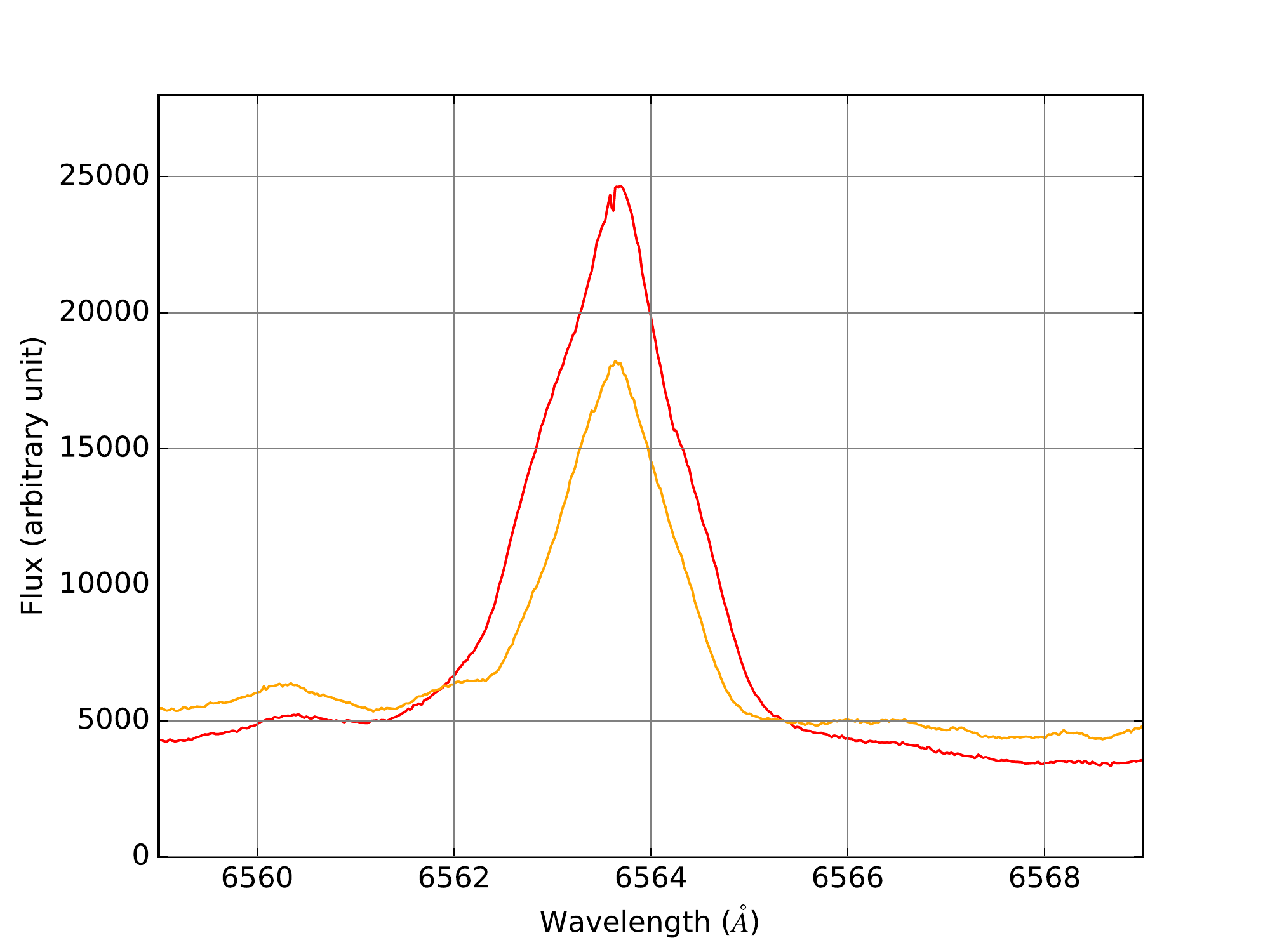}
        \caption{Region around the He\,I emission line at $5875\,\AA$ and the Na\,I doublet at $5890-5896\,\AA$ (top panel) in the UVES spectra of GJ65 A (orange) and B (red). The very intense H$\alpha$ emission line is shown in the bottom panel. The spectra are not corrected for telluric lines. \label{uves_emission}}
\end{figure}

We retrieved from the ESO archive the observation of GJ65 A and B obtained using the VLT/UVES spectrograph \citepads{2000SPIE.4008..534D}\footnote{Details on the instrument can be found at \url{http://www.eso.org/sci/facilities/paranal/instruments/uves.html}} on 17 November 2005 between UT00:39 and UT01:43.
These observations are particularly interesting as the two stars were aligned on the entrance slit, thus providing separate spectra of the two stars.
According to the DIMM, the seeing was around $1.0\arcsec$ in the $V$ band, resulting in very little crosstalk between the two echelle spectra on the detector. The airmass was comprised between 1.13 ans 1.08 throughout the exposure sequence.
The raw data were processed with version 5.7.0 of the UVES data pipeline\footnote{Available from \url{http://www.eso.org/sci/software/pipelines/}} \citepads{2000Msngr.101...31B, uvesmanual}, and using the associated \emph{Reflex} workflow \citepads{2013A&A...559A..96F}.
We focused on the wavelength range between 4800 and $6800\,\AA$.
This limits the effect of the poorer seeing at shorter wavelengths and of telluric lines at longer wavelengths. An overview of the two spectra is presented in Fig.~\ref{uves_global}.
Prominent emission lines are visible in the spectra of the two stars, in particular in the region of the He\,I line and Na\,I doublet around $5890\,\AA$ and in the H$\alpha$ line (Fig.~\ref{uves_emission}).
The fainter star GJ65\,B exhibits systematically more intense emission lines than GJ65\,A, indicating a stronger magnetic chromospheric activity (see also Sect.~\ref{starspots}). Similar emission lines were observed during a flare of the M4V dwarf \object{GJ699} by \citetads{2006PASP..118..227P}.

\section{Properties of GJ65 A \& B}\label{analysis}

\subsection{Flaring and interferometric closure phases \label{t3phi-discussion}}

The strong and persistent surface magnetic fields of active M dwarfs produces frequent and remarkably violent flares, lasting up to several hours and reaching high luminosities.
Ultrafast microflaring has recently been detected at optical wavelengths by \citetads{2016A&A...589A..48S}.
The flares sometime approach the star's bolometric luminosity \citepads{2010ApJ...714L..98K}. Owing to its proximity and particularly high level of activity, GJ65 has been intensively studied for decades over a broad range of wavelengths, from X-rays (see, e.g., \citeads{2014ATel.6026....1I}) to the radio domain.
\citetads{1974ApJS...29....1M} established observationally that UV\,Ceti shows on average one flare every 36\,minutes. The linear separation of the two stars varies between approximately 2 and 8\,AU during their orbital cycle. Although it was proposed by \citetads{2011Ap.....54..469M} that the frequency of the flares increases (from 0.2 to 2 flares/hour) when the distance between the two stars decreases, the smallest separation of the two stars is still too large to realistically expect a significant magnetic or photometric interaction between them.

Our PIONIER observations allow us to search for the signature of the radio flares in the near-infrared, using the closure phase signal as a proxy for the degree of departure of the star from central symmetry.
We did not detect any significant closure phase signal in our PIONIER data.
This non-detection is consistent with the prediction by \citetads{2010ApJ...714L..98K} that the flaring activity should result in very small photometric effects in the infrared, on the order of 0.01\,mag. \citetads{2012ApJ...748...58D} extended the ultraviolet-visible modeling of flares to the near-IR domain and concluded that even an exceptional flaring event causing a ultraviolet flux increase of $\Delta u=5$\,mag would result in a flux increase of only a few 0.01\,mag in the near-infrared. This surprisingly small contribution in the infrared is due to the combination of the intrinsic blue color of the flares, and the very red color of the M dwarf spectrum. Their influence on the PIONIER data is therefore too limited to be detectable.

\subsection{Differential radial velocity $V_\mathrm{rad}(B) - V_\mathrm{rad}(A)$\label{radvel}}

\begin{figure*}[]
        \centering
        \includegraphics[width=15cm]{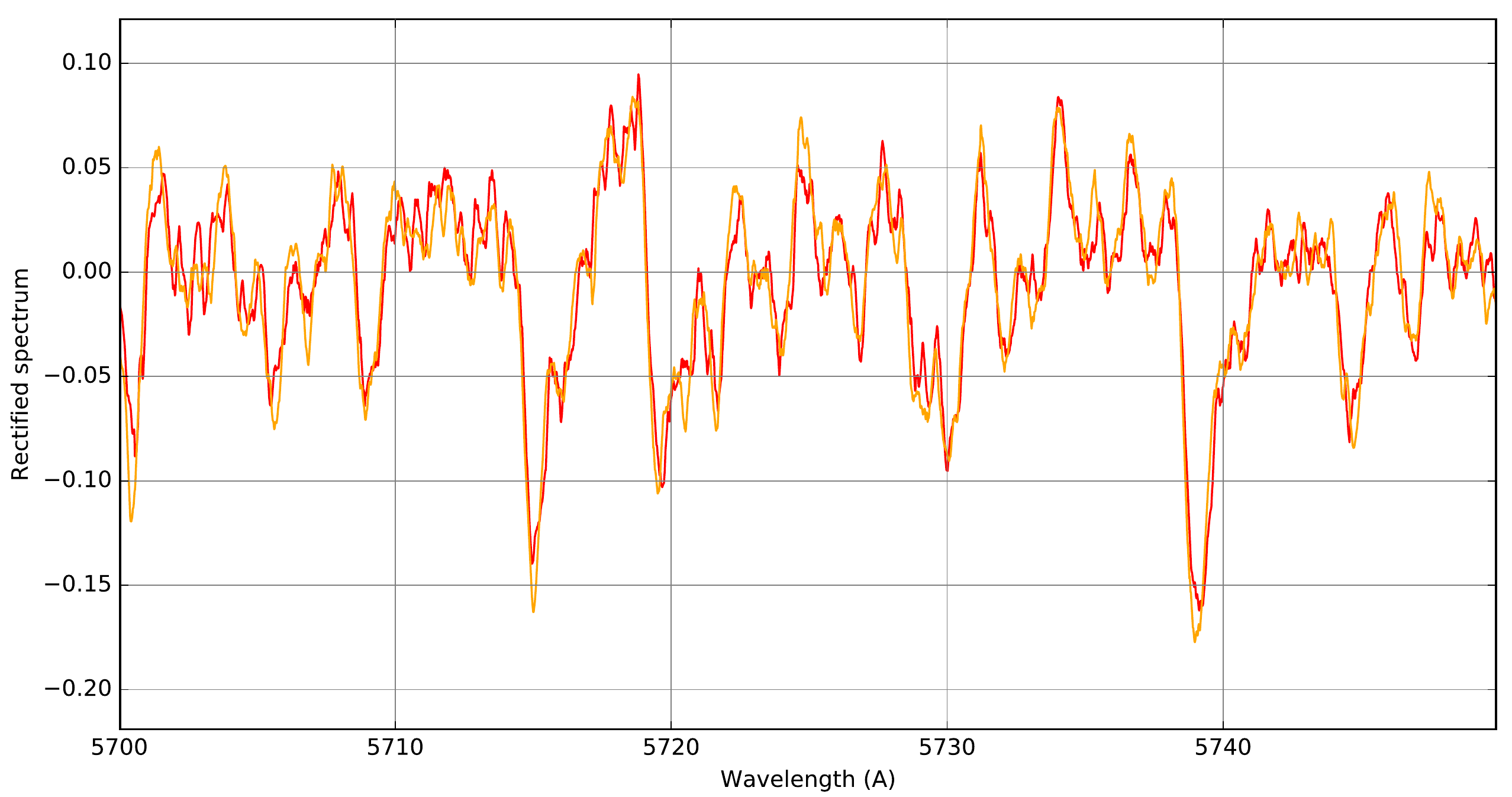}
        \caption{Rectified spectra of GJ65 A (orange) and B (red), before Doppler-shift recentering. The vertical scale is the relative deviation in flux with respect to the moving median of each spectrum over a $5\,\AA$ window.\label{rectified_spectra}}
\end{figure*}

Using our UVES spectra (Sect.~\ref{uves_data}), we measured the difference in radial velocity between the two stars to use it to constrain the orbit of the system.

We first high-pass-filtered the reduced spectra (Fig.~\ref{uves_global}) by dividing them by a smoothed version of themselves using a moving median with a broad window of 5\,nm (300 spectral channels of 17\,pm). We then subtracted one from the result to bring them to zero average.
This normalization allowed us to calibrate the slightly different color and flux level of the two stars.
The main advantage of high-pass-filtered spectra (hereafter ``rectified spectra'') is that they can be easily compared between different stars, as they are flat and exhibit comparable line amplitudes (expressed in fractions of the moving median), symmetrically around zero.
The flux from GJ65\,A was found to be approximately 40\% higher than the flux of B on average over the spectrum, and slightly redder. This may be due to an ongoing flare event at the time of the observation, which would make GJ65\,B appear bluer than in its quiescent state.
We derived the relative radial velocity offset between the two stars by adjusting $\Delta V_\mathrm{rad} = V_\mathrm{rad}(B) - V_\mathrm{rad}(A)$ to minimize the standard deviation of the difference of the two shifted spectra. 
For the fit, we symmetrically shifted the two spectra by $\Delta V_\mathrm{rad}/2$ and $-\Delta V_\mathrm{rad}/2$ for GJ65\,A and B, respectively. This approach resulted in the same resampling of the spectra, providing a better match of the noise level and equal weight of the two spectra in the fit.

We obtain a radial velocity shift of $\Delta V_\mathrm{rad} = +2.323 \pm 0.068\,\mathrm{km\,s}^{-1}$ for epoch 17 November 2005, UT00:54 ($\mathrm{MJD} = 53691.0375$). The positive sign indicates that the spectrum of GJ65\,B is redshifted compared to GJ65\,A.
The uncertainty on this measurement is taken as the scatter of the computed Doppler velocities over 18 separate spectral segments of $\delta\lambda = 20$ to $50\,\AA$ in width, covering about $800\,\AA$ in total and spread over $\lambda = 5200$ to $6750\,\AA$.
Our accuracy on $\Delta V_\mathrm{rad}$ is very good thanks to the high degree of similarity between the spectra of the two stars.
The spectral segments used in the fit were selected to match
well between the rectified spectra of A and B. We avoided the emission line regions, which have different line widths and intensities in the two stars. An illustration of the good correspondence of the spectra is presented in Fig.~\ref{rectified_spectra}. 
The residual dispersion after recentering the spectra is on average only 2\% of the flux over the 18 selected spectral regions, indicative of a good quality of the fit.

\subsection{Projected rotational velocities, heliocentric radial velocities, and metallicity\label{rotvel}}
We estimated the rotational velocities of GJ65\,A \& B by matching their rectified spectrum with a reference spectrum convoluted by a theoretical rotation profile \citepads[see, e.g.,][]{2009ApJ...704..975J}.

As the fiducial star, we selected Proxima, which has been extensively observed with UVES.
This red dwarf has a spectral type very similar
to that of GJ65 A and B, but it is a slow rotator, with $v \sin i = 2$\,km\,s$^{-1}$ \citepads{2014MNRAS.439.3094B}, and therefore a good template spectrum for the rotational velocity fit.
We retrieved from the ESO Phase 3 Science Archive a UVES spectrum of Proxima obtained on 24 January 2005 UT08:21\footnote{File reference: {\tt ADP.2013-12-06T08:42:53.873}} ($\mathrm{MJD} = 53394.348$).
The mathematical expression of the rotation profile was defined following the formalism of \citetads{2011A&A...531A.143D}.
For the limb darkening, we adopted the linear coefficient $u_R = 0.8557$ derived by \citetads{2012A&A...546A..14C} for the $R$ band (flux conservation, same model parameters as in Sect.~\ref{pionier_obs}).
We minimized the standard deviation of the residuals of the fit of the spectrum of Proxima on those of GJ65 A and B by simultaneously varying the projected rotational velocity $v \sin i$ and the line depth ratio $\beta$ defined as a multiplicative factor applied to the spectrum of Proxima.
Within this process, we also fit the relative radial velocities of GJ65 A and B with respect to Proxima. We adopted as fiducial reference the heliocentric radial velocity of $V_\mathrm{rad}(Proxima) = -22.345 \pm 0.006$\,km\,s$^{-1}$ determined by \citetads{2014MNRAS.439.3094B}.
We obtain heliocentric velocities of  $V_\mathrm{rad}(A) = +37.83 \pm 0.20$\,km\,s$^{-1}$ and $V_\mathrm{rad}(B) = +40.29 \pm 0.20$\,km\,s$^{-1}$ , giving a relative radial velocity $\Delta V_\mathrm{rad}(B-A) =  +2.45 \pm 0.20$\,km\,s$^{-1}$.
The uncertainties were derived by varying the adjustment parameters (wavelength range, smoothing window, etc.) and are reasonably conservative taking into account the long-term stability of UVES \citepads{2000SPIE.4005..121D}.
The differential velocity $\Delta V_\mathrm{rad}$ agrees well with the value we derived in Sect.~\ref{radvel} by matching the spectra of GJ65\,A and B (i.e.,~not using Proxima's spectrum).
These results are also close to the combined radial velocity for GJ65 A and B of $V_\mathrm{rad}(AB) = +42.4$\,km\,s$^{-1}$ listed in the Palomar/MSU survey \citepads{1995AJ....110.1838R}.

We obtain the following $v \sin i$ and $\beta$ of the components of GJ65: $v \sin i (A) = 28.2 \pm 2$\,km\,s$^{-1}$, $\beta(A) = 0.84 \pm 0.05$,  $v \sin i (B) = 30.6 \pm 2$\,km\,s$^{-1}$ , and $\beta(B) = 0.67 \pm 0.05$. The projected rotational velocities agree very well with the values of $v \sin i (A) = 31.5 \pm 3$\,km\,s$^{-1}$ and  $v \sin i (B) = 29.5 \pm 3$\,km\,s$^{-1}$ published by \citetads{2005MNRAS.358..105J}.

\begin{figure}[]
        \centering
        \includegraphics[width=\hsize]{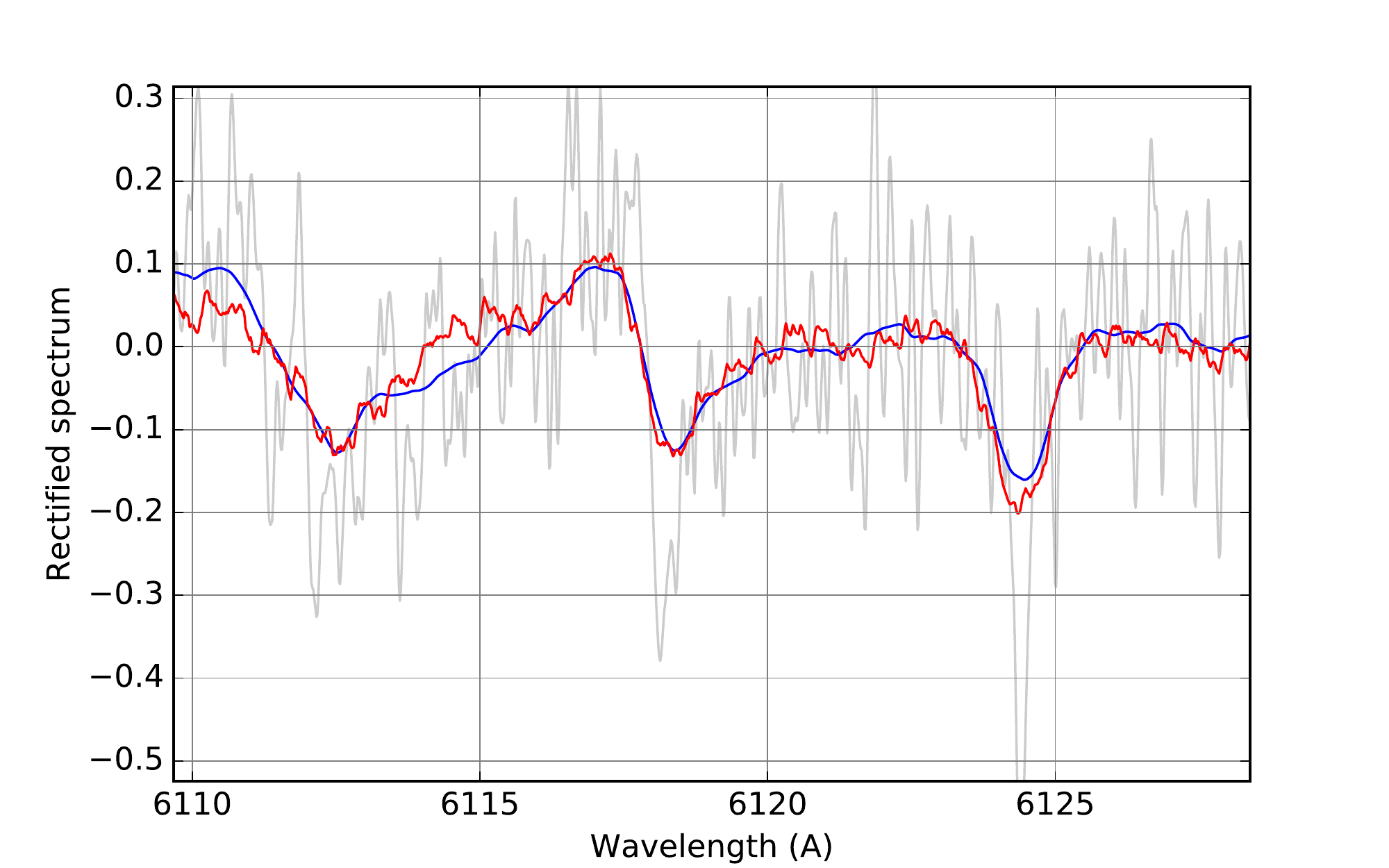}
        \caption{Enlargement of the rectified spectra of GJ65 B (red) and Proxima before (light gray) and after (blue) convolution by a rotation profile with $v \sin i = 31$\,km\,s$^{-1}$ and multiplication by a line depth correction factor $\beta = 0.67$.\label{rotation_fit}}
\end{figure}

We interpret the values of $\beta$ as a proxy for the relative metallicity of GJ65 A and B with respect to Proxima. Assuming that the curves of growth of the lines present in the spectrum do not saturate, we can convert the value of $\beta$ into an additive number for the metallicity through $\Delta [\mathrm{Fe/H}] = \log(\beta)$.
The metallicity of Proxima is close to solar, as \citetads{2016A&A...587A..19P} found $[\mathrm{Fe/H}] (\mathrm{Proxima)} = -0.07 \pm 0.14$ and \citetads{2014A&A...568A.121N} $+0.16 \pm 0.20$.
Adopting a value of $+0.05 \pm 0.20$\,dex for Proxima, our simple differential approach gives $[\mathrm{Fe/H}](\mathrm{A}) = -0.03 \pm 0.20$\,dex and $[\mathrm{Fe/H}](\mathrm{B}) = -0.12 \pm 0.20$\,dex.
According to \citetads{2012A&A...538A.143K}, the metallicity of GJ65 is sub-solar at ($[\mathrm{Fe/H}]=-0.42 \pm 0.10$), which is at 2$\sigma$ from our mean value.
But this value is derived from model fitting on a low-resolution combined spectrum of A and B ($R=1000$, $\Delta v \approx 160$\,km\,s$^-1$), which likely biases $[\mathrm{Fe/H}]$ toward lower values for these fast-rotating stars.
For this reason, we argue that GJ65 has in fact a metallicity close to solar.

From the separated UVES spectra presented in Sect.~\ref{uves_data}, complemented by the $K$ band high-resolution spectrum ($R=18\,000$) of GJ65\,B from the Gemini/GNIRS spectral library\footnote{\url{http://www.gemini.edu/sciops/instruments/nearir-resources/spectral-templates?q=node/11594}} \citepads{2009ApJS..185..186W}, it is possible to determine the metallicity of each component of GJ65 for individual chemical element.
This is done by comparing the observed spectra with recent model atmospheres, as listed~by \citetads{2012RSPTA.370.2765A}\footnote{\url{https://phoenix.ens-lyon.fr/simulator/index.faces}}
, for example. This effort is beyond the scope of the present work.

\subsection{Orbital parameters and masses\label{orbital-solution}}

We consider our new astrometric measurements of the separation $\rho$ and position angle $\theta$ (with respect to North) of GJ65\,AB listed in Table~\ref{table-astrom}, complemented by the archival astrometry listed by \citetads{1988AJ.....95.1841G} and additional measurements from the Washington Double Star catalog \citepads{2016yCat....102026M, 2001AJ....122.3466M}.
The error bars associated with the pre-1990 data are generally not available, and we chose the following uncertainties: $0.25\arcsec$ before 1935, $0.15\arcsec$ between 1925 and 1985, $0.1\arcsec$ between 1985 and 1990, and $0.05\arcsec$ for more recent data. The uncertainties of the derived parameters were normalized to the observed data dispersion.
We also included in our orbit determination our single UVES differential radial velocity measurement reported in Sect.~\ref{radvel}.
The combination of the high-accuracy parallax with the HST and adaptive optics differential astrometry results in a much improved estimate of the orbital parameters and masses of the two stars (Fig.~\ref{orbit-fit}).
The derived orbital elements are listed in Table~\ref{orbit-elements}, where we also summarize the physical parameters of the system from the present work and the literature.
The reduced $\chi^2$ of the fit is 1.10, indicative of a good fidelity of the determined orbit with respect to the observations. No significant residual signal is observed on the recent high-accuracy differential astrometry, which has a dominant weight in the fit.
The total mass of the system is estimated to be $m_\mathrm{tot} = \Mtot \pm \Mtoterr\,M_\odot$. The contributions to the error budget of the total mass are $\sigma(\pi)=0.005\,M_\odot$, $\sigma(a) = 0.003\,M_\odot$, and $\sigma(P) = 0.001\,M_\odot$.

\begin{figure}[]
        \centering
        \includegraphics[width=8cm]{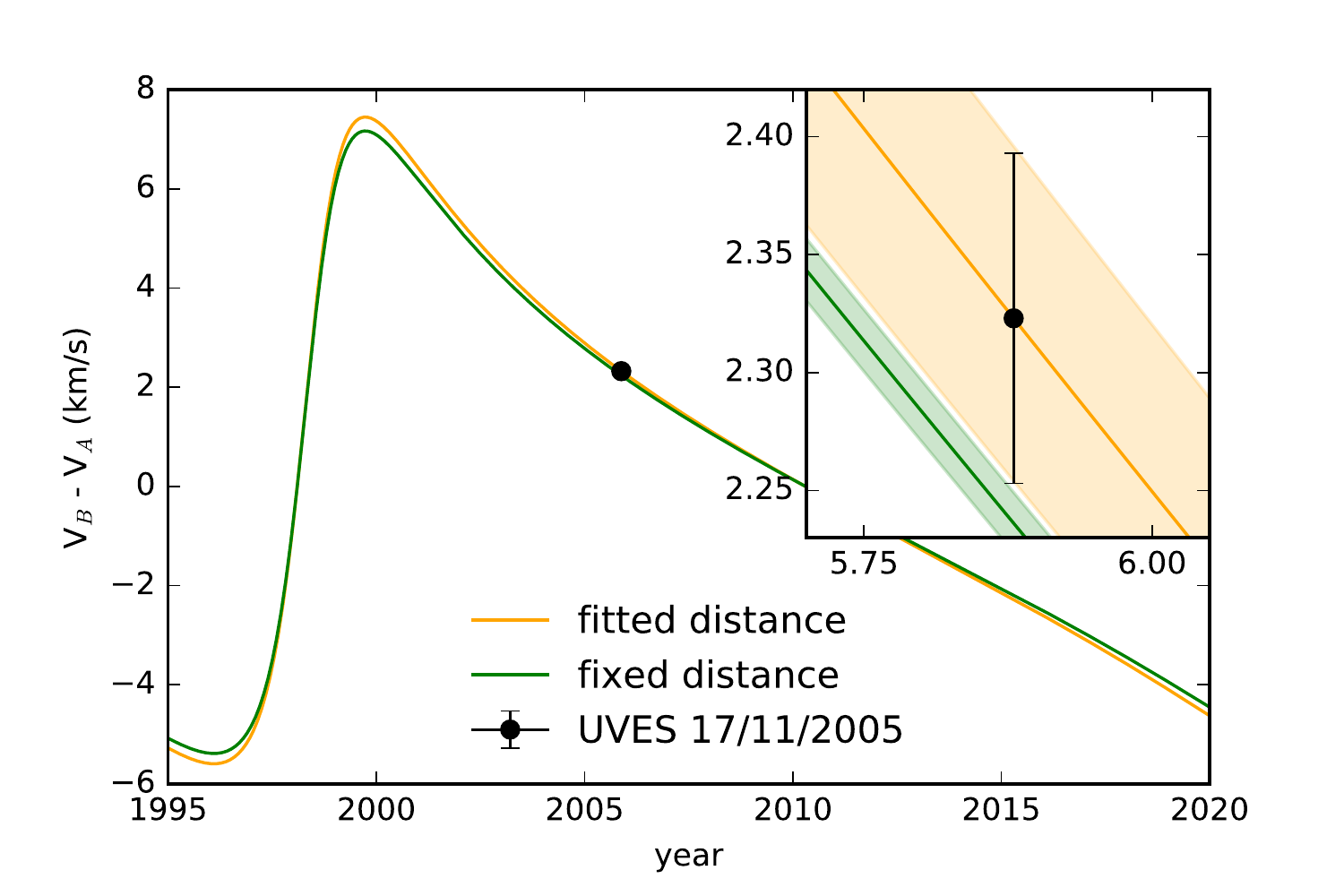}
        \caption{Prediction of the differential radial velocity $\Delta V_\mathrm{rad} = V_\mathrm{rad}(B) - V_\mathrm{rad}(A)$ from orbital models constrained without the UVES measurement as a constraint (green curve) and including it, but considering the parallax as a free parameter (orange curve). The model uncertainty is represented as the shaded areas. \label{orbit-uves}}
\end{figure}

\begin{figure*}[]
        \centering
        \includegraphics[width=16cm]{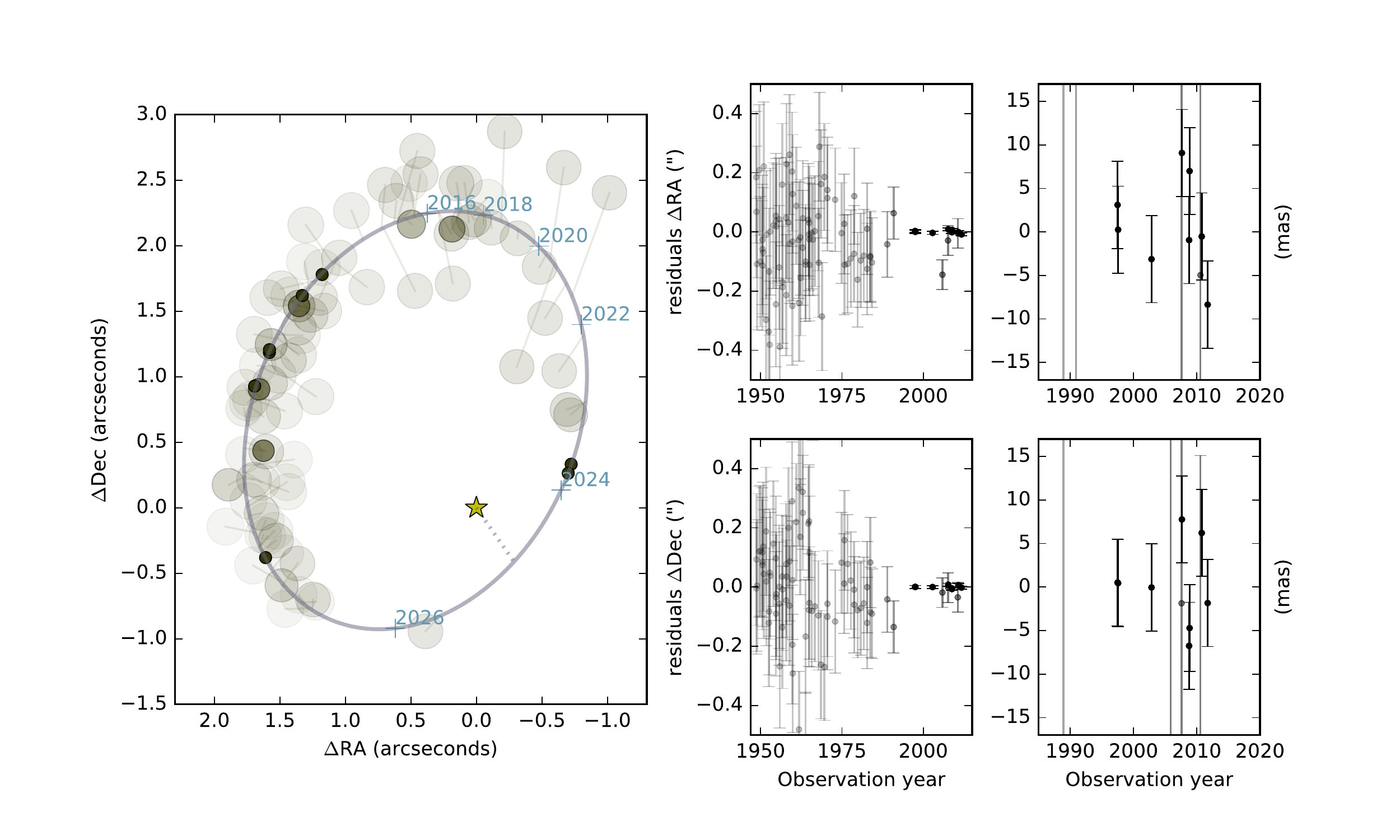}
        \caption{Left: orbital fit of GJ65\,AB (left). The measurements are shaded according to their accuracy (darker for more accurate measurements). Right: residuals of the fit, with an enlargement of the recent high-accuracy measurement epochs. \label{orbit-fit}}
\end{figure*}

The total mass of the system is affected only marginally (at the $0.001\,M_\odot$ level) when we remove the UVES differential radial velocity from the fit. This is due to the good match of the measured $\Delta V_\mathrm{rad}$ compared to the prediction of the model constrained without this point (Fig.~\ref{orbit-uves}, green curve).
We also tested an orbital fit including the UVES points and with the parallax $\pi$ as a free parameter (Fig.~\ref{orbit-uves}, orange curve), from which we obtain $\pi = 359 \pm 12$\,mas and a total mass $m_\mathrm{tot} = 0.27 \pm 0.03\,M_\odot$. Both $\pi$ and $m_\mathrm{tot}$, although less precise, remain statistically compatible with the best-fit values with all constraints ($\Delta \pi = -1.2\sigma$, $\Delta M = +0.9\sigma$).
Forcing the fit to pass through the UVES measurement while keeping the parallax by \citetads{1995gcts.book.....V} as a constraint results in a total mass of $0.244\,M_\odot$ , which is well within our uncertainty range.
The inclusion of $\Delta V_\mathrm{rad}$ in the fit allows us to solve the ambiguity on the inclination $i$ of the orbit (Table~\ref{orbit-elements}).

The total mass we find is compatible within the combined error bars with the value of $0.224 \pm 0.016\,M_\odot$ measured by \citetads{1973AJ.....78.1096H}, but it is higher by 20\%\ than the total mass of $0.200\,M_\odot$ derived by \citetads{1988AJ.....95.1841G}. This difference probably arises because only half of the astrometric orbit was measured with a relatively good accuracy at the time of the latter publication (larger shaded points in the left panel of Fig.~\ref{orbit-fit}).
Determining the mass ratio $m_B/m_\mathrm{tot}$ of GJ65 from photographic plate astrometry is a relatively delicate task because the stars are unresolved in the seeing-limited regime for part of their orbit. This results in a dependence of the differential astrometry on the magnitude difference of the two stars during the unresolved section of the orbit. However, \citetads{1973AJ.....78.1096H} established that the dependence on the magnitude difference is small, and therefore that this measurement is essentially geometrical.
For this reason, we adopted the mass ratio determined by \citetads{1988AJ.....95.1841G}, $m_B/m_\mathrm{tot} = 0.4938 \pm 0.0031$, which agrees well with \citetads{1973AJ.....78.1096H} and \citetads{1987AJ.....94.1077H}. Using this value, we obtain individual masses of $m_A = \MA \pm \MAerr\,M_\odot$ and $m_B = \MB \pm \MBerr\,M_\odot$.
As a remark, the contribution of the mass ratio uncertainty ($\pm 0.6\%$) to our error bars on the individual masses ($3.6\%$) is negligible. Adopting the $0.502 \pm 0.02$ mass ratio of \citetads{1987AJ.....94.1077H},
for instance, leads to a change of only $1.6\%$ of the individual masses, which is less than half our error bar.

Applying the mass-luminosity (M--L) relations calibrated by \citetads{2000A&A...364..217D} to the absolute near-infrared magnitudes listed in Table~\ref{table-magdiff} results in predicted masses of $m_A = 0.109\,M_\odot$ and $m_B = 0.102\,M_\odot$, which is 12\% and 17\% lower than the measured masses, respectively. The reason may be that the absolute magnitudes of GJ65\,AB are close to the limit of applicability of the M--L relations (11, 10, and 9.5 for the $JHK$ absolute magnitudes,
respectively).

\begin{table}
        \caption{Summary of the orbital and physical parameters of GJ65 A and B. The adjusted data set includes the available astrometry of our UVES differential radial velocity measurement.}
        \centering          
        \label{orbit-elements}
        \begin{tabular}{lrl}
        \hline\hline
        \noalign{\smallskip}
        Parameter & Value & Ref.\\
        \noalign{\smallskip}
        \hline    
        \noalign{\smallskip}
        Parallax $\pi$ (mas) & $373.70 \pm 2.70$ & (c) \\
        \hline    
        \noalign{\smallskip}
        {\it Orbital parameters:} \\
        $\Omega$ (deg) & $145.79 \pm 0.32$ & (a) \\
        Period $P$ (years) &  $26.284 \pm 0.038$ & (a) \\
        T$_0$ (year) &  $1972.115 \pm 0.047$ & (a) \\
        $a$ (arcseconds) &   $2.0584 \pm 0.0097$ & (a) \\
        Eccentricity $e$ &     $0.6185 \pm 0.0020$ & (a) \\
        Inclination $i$ (deg) &   $307.82 \pm 0.28$ & (a) \\
        $\omega$ (deg) & $283.27 \pm 0.21$ & (a) \\
        \hline    
        \noalign{\smallskip}
        {\it Masses:}\\
        $m_\mathrm{tot} = m_A + m_B$ ($M_\odot$) & $\Mtot \pm \Mtoterr$ & (a) \\
        $m_B / m_\mathrm{tot}$ & $0.4938 \pm 0.0031$ & (b) \\
        $m_A$ ($M_\odot$) & $\MA \pm \MAerr$ & (a+b) \\
        $m_B$ ($M_\odot$) & $\MB \pm \MBerr$ & (a+b) \\
        \hline    
        \noalign{\smallskip}
        {\it Radii:}\\
        $R_A$ ($R_\odot$) & $\radA \pm \radAerr$ & (a+c) \\
        $R_B$ ($R_\odot$) & $\radB \pm \radBerr$ & (a+c) \\
        \hline    
        \noalign{\smallskip}
        {\it Surface gravity:}\\
        $\log g_A$ (cgs) & \loggA & (a+b+c) \\
        $\log g_B$ (cgs) & \loggB & (a+b+c) \\
        \hline    
        \noalign{\smallskip}
        {\it Rotational $v\,\sin i $:}\\
        A (km/s) & $28.2 \pm 2$ & (a) \\
        B (km/s) & $30.6 \pm 2$ & (a) \\
        \hline    
        \noalign{\smallskip}
        {\it Metallicity:} \\
        $[\mathrm{Fe/H}](A)$ (dex) & $-0.03 \pm 0.20$ & (a) \\
        $[\mathrm{Fe/H}](B)$ (dex) & $-0.12 \pm 0.20$ & (a) \\
        \hline    
        \noalign{\smallskip}
        {\it Radial velocities:} & (MJD=53691.0375)\\
        $\Delta V_\mathrm{rad}(B-A)$ (km/s) & $+2.323 \pm 0.068$ & (a) \\
        $V_\mathrm{rad}(A)$ (km/s) & $+37.83 \pm 0.20$ & (a) \\
        $V_\mathrm{rad}(B)$ (km/s) & $+40.29 \pm 0.20$ & (a) \\
        Barycentric $V_\mathrm{rad}$ (km/s) & $+39.04 \pm 0.20$ & (a+b) \\
        \hline    
        \noalign{\smallskip}
        {\it Galactic space velocity:}\\
        $(U,V,W)$ (km/s, LSR) & (-34.93, -6.88, -21.73) & (a+d) \\
        \hline                      
        \end{tabular}
        \tablefoot{
        (a) present work;
        (b) \citetads{1988AJ.....95.1841G};
        (c) \citetads{1995gcts.book.....V};
        (d) \citetads{2003ApJ...582.1011S}.
        }
\end{table}

\subsection{Limits on low-mass companions of GJ65}

We used the CANDID code \citepads{2015A&A...579A..68G} to search for companions in the close vicinity of the components of GJ65 using the PIONIER squared visibilities and close phases (Sect.~\ref{pionier_obs}). We did not detect any faint companion down to the following $H$ -band magnitude contrasts and angular separations $\rho$:
GJ65\,A: $\Delta H_A = 4.0$ for $\rho_A = 3$-10\,mas, $\Delta H_A = 4.4$ for $\rho_A = 10$-30\,mas; GJ65\,B: $\Delta H_B = 3.6$ for $\rho_B = 3$-10\,mas, $\Delta H_B = 4.0$ for $\rho_B = 10$-30\,mas.
Considering the absolute magnitudes of A and B listed in Table~\ref{table-magdiff} , this corresponds to absolute magnitude detection limits around $M_H = 13.5$. According to the AMES-Cond models \citepads{2003A&A...402..701B}, this corresponds to a mass of about $50\,M_J$ at an age of 1\,Gyr, and $70\,M_J$ at 5\,Gyr.

We did not detect any source other than GJ65 A and B in our NACO images in the $JHK_sL'$ bands (Sect.~\ref{naco_obs}) in a field of view of several arcseconds. However, we did not conduct a thorough analysis of the sensitivity limits.

Our orbital fit does not show astrometric residuals in the period 1997-2011 at a level of approximately $\pm 5$\,mas. This theoretically limits the possible presence of a massive companion orbiting either of the individual dwarf stars at a level of approximately $0.6 / a_P$ Jupiter masses, with $a_P$ the semi-major axis of the planet orbit in arcseconds.
Considering the size of the orbit of the GJ65 pair, setting an upper limit of $0.2\arcsec$ to the potential planet orbital separation around either of the two stars results in a sensitivity limit of a few Jupiter masses. Our time coverage is very insufficient to conclude at this level of sensitivity, however, because a regular astrometric coverage at milliarcsecond accuracy over several years would be necessary.
Our non-detection of companions is in line with the conclusion of \citetads{2009ApJ...701.1922B} from radio astrometry of GJ65\,B.

\section{Comparison with Proxima\label{proxima}}

\subsection{GJ65\,AB age, metallicity, and the mass-radius relation \label{mass-radius}}
 
\begin{figure}[]
        \centering
        \includegraphics[width=\hsize]{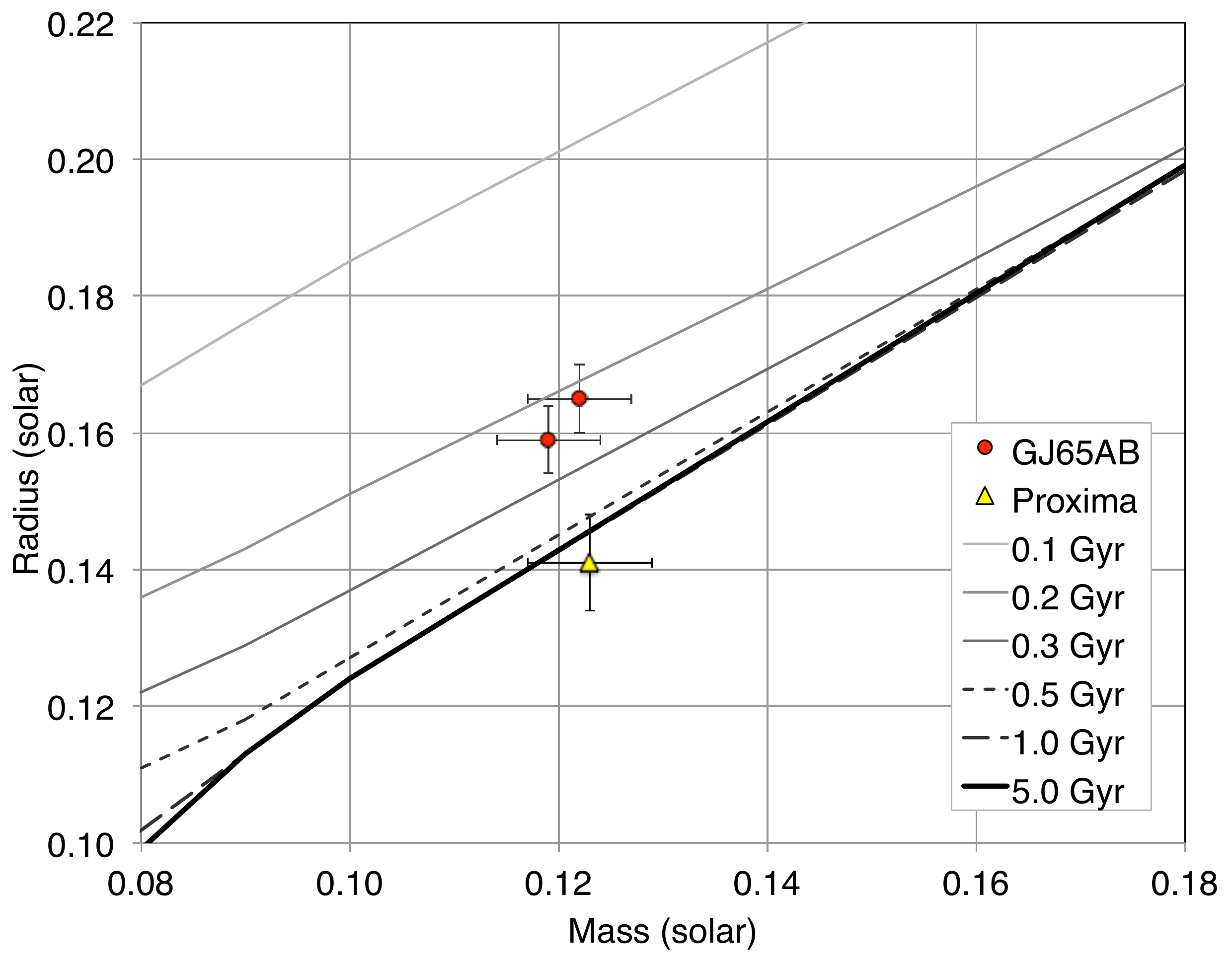}
        \includegraphics[width=\hsize]{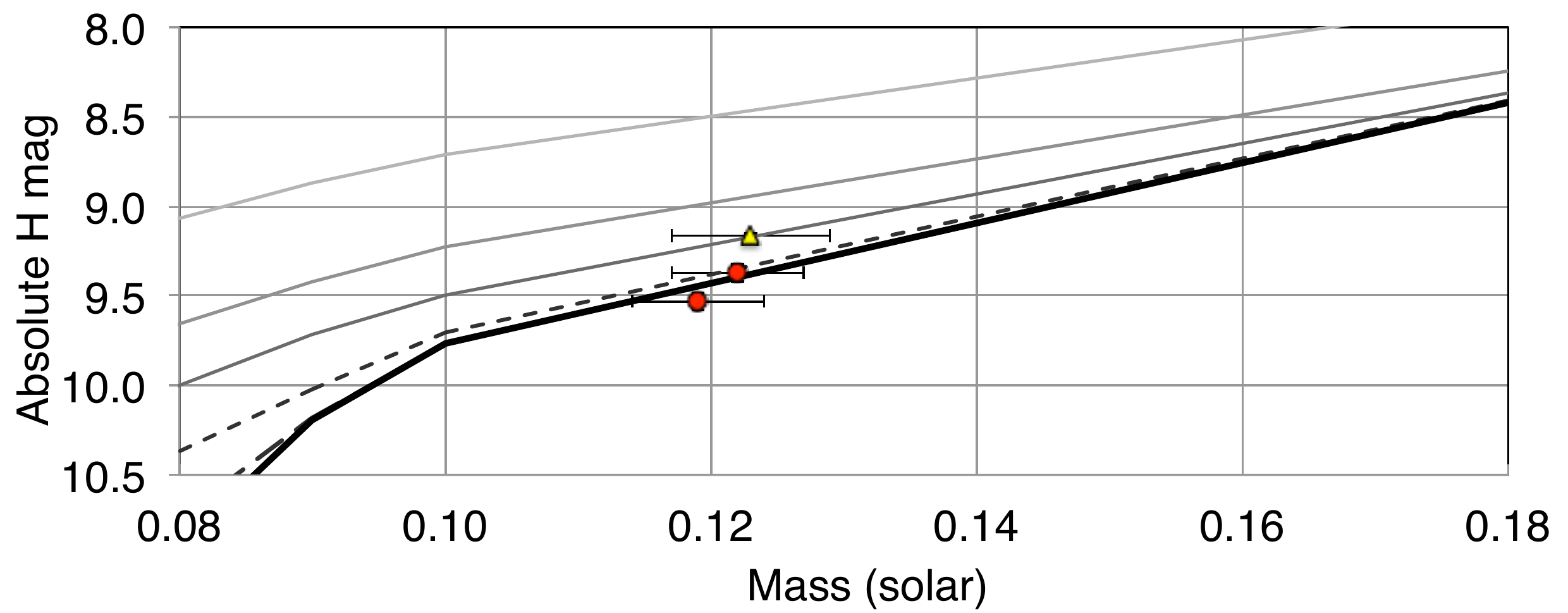}
        \caption{Positions of GJ65 A and B (red points) and Proxima (yellow triangle) in the mass-radius (top) and mass-absolute $H$ magnitude (bottom) diagrams with \citetads{2015A&A...577A..42B} isochrones.
        \label{mr-diagram}}
\end{figure}

In Fig.~\ref{mr-diagram} (top panel) the positions of \object{GJ65} AB and \object{Proxima} in the mass-radius (M--R) diagram are superimposed on theoretical isochrones from \citetads{2015A&A...577A..42B}, hereafter BHAC15. 
In this diagram, the radii of GJ65 A and B are consistent with very young stars aged between 200 and 300\,Myr.
However, such a young age is incompatible with the observed absolute infrared magnitudes of GJ65 A and B, which are too faint compared to young BHAC15 models by approximately 0.3\,mag in the $JHK$ bands. This is illustrated in Fig.~\ref{mr-diagram} (bottom panel) for the $H$ band, but the same behavior is observed in the $J$ and $K$ bands.

To test whether GJ65 belongs to the old population of the Galactic thick disk, we determine in this paragraph its Galactic space velocity vector.
The proper motion of GJ65 is $\mu_\alpha = +3296.2 \pm 5.5$\,mas\ yr$^{-1}$
and $\mu_\delta = +563.9 \pm 5.5$\,mas\ yr$^{-1}$ \citepads{2003ApJ...582.1011S}.
The mass ratio $m_B/m_\mathrm{tot}$ together with the radial velocities of GJ65 A and B give a barycentric radial velocity of $V_\mathrm{rad} = +39.04 \pm 0.20$\,km~s$^{-1}$.
From its coordinates, parallax, proper motion and barycentric radial velocity, the Galactic space velocity vector of GJ65
in the local standard of rest (LSR) is $(U, V, W) = (-34.93, -6.88, -21.73)$~km~s$^{-1}$, which is consistent with \citetads{1995ahsm.book.....C}.
We considered for this computation the J2000 transformation matrix to Galactic coordinates from the introduction to the \emph{Hipparcos} catalog, and we assumed a velocity of the Sun with respect to the LSR of $(U,V,W)_\odot = (+11.1, +12.2, +7.25)$~km~s$^{-1}$ 
\citepads{2010MNRAS.403.1829S}.
We followed the convention of $U$ positive toward the Galactic center, $V$ in the direction of Galactic rotation, and
$W$ toward the North Galactic pole.
According to Fig.~7 of \citetads{2011MNRAS.412.1237C}, this velocity vector indicates that GJ65 probably belongs to the thin disk of the Milky Way, so the determined Galactic space velocity vector does not constrain significantly its age. The velocity vector components however indicate that the binary is probably a member of the old thin disk population, with an age between 1 and 8\,Gyr (Croswell, private communication).

The radii we derive for GJ65 A and B are $\approx 15\%$ larger than the radius of \object{Proxima} (\object{GJ551}): $R(\mathrm{Proxima}) = 0.141 \pm 0.007\,R_\odot$ \citepads{2009A&A...505..205D}. The masses are almost identical for the three stars as $M(\mathrm{Proxima}) = 0.123 \pm 0.006\,M_\odot$, which is within 1$\sigma$ of our determinations of the masses of both GJ65 A and B (Table~\ref{orbit-elements}).
In contrast with GJ65 A and B, the position of Proxima in the M--R diagram is well reproduced by the BHAC15 models. Its mass was predicted by \citetads{2009A&A...505..205D} from the M--L relations by \citetads{2000A&A...364..217D}. As discussed in Sect.~\ref{naco_obs}, the accuracy of these relations may not be very good for such VLMS whose absolute magnitudes are close to the limit of the calibrated range. However, the mass of Proxima would have to be unrealistically low ($\approx 0.09\,M_\odot$) to differ from the BHAC15 models as much as GJ65 A and B.
The agreement for Proxima is also better on the effective temperature: \citetads{2003A&A...397L...5S} obtained T$_\mathrm{eff} = 3042 \pm 117$\,K, close to the value of T$_\mathrm{eff} = 3054 \pm 79$\,K by \citetads{2012ApJ...757..112B} listed in the PASTEL catalog \citepads{2010A&A...515A.111S}. This is compatible within 1$\sigma$ with the BHAC15 prediction of 2900\,K.

In summary, GJ65 A and B have infrared absolute magnitudes consistent with an age of at least 0.5 to 1\,Gyr, possibly even much older. But their radii are above the 1\,Gyr and 5\,Gyr isochrones from BHAC15 (that are mostly identical) by $\left( R_\mathrm{obs} - R_\mathrm{model} \right) / R_\mathrm{model} = +14 \pm 4\%$ and $+12 \pm 4\%, $ respectively.

\subsection{Metallicity \label{metal}}
The low metallicity of GJ65 found by \citetads{2012A&A...538A.143K} compared to Proxima may be invoked as a reason for the radius discrepancy. 
However, we obtain metallicity values very close to solar from our simple differential analysis of the UVES spectra of GJ65\,AB and Proxima.
On the theoretical side, the possibility that metallicity influences significantly the radius was studied by \citetads{2013ApJ...776...87S} for a sample of low-metallicity and low-mass stars ($M \lesssim 0.42\,M_\odot$) with interferometric radius measurements, but they concluded that a change in model metallicity is not sufficient to reconcile the models and observations.
For these reasons, we argue that the metallicity difference between GJ65 and Proxima, if it exists, is insufficient to explain the radius difference between these stars.

\subsection{Equation of state}
\citetads{2015A&A...577A..42B} pointed out that the M--R relationship for VLMS is defined essentially by the equation of state (EOS) of the internal matter (which is partially degenerate) and not by the atmospheric properties.
The difference that we find between our observations and the BHAC15 models for GJ65\,AB could thus in principle be explained by an inaccurate definition of the EOS.
However, the good agreement of the predicted radius for Proxima points at a reasonably good definition of the EOS. 
As the masses, metallicity, and internal physical conditions are very similar for the three stars, a significant error in the EOS parameters would result in a common discrepancy with respect to the M--R relation for all three stars. We therefore believe that a significant error in the EOS is unlikely.

\subsection{Rotational velocity}
\citetads{2007A&A...472L..17C} proposed that the efficiency of the convective energy transport in fast-rotating magnetic VLMS is reduced compared to single stars.
Comparing GJ65\,AB and Proxima, this appears as a reasonable explanation, as the principal physical difference between these three stars is their rotation velocity. While $v \sin i \approx 30$\,km/s for both components of \object{GJ65} as found by \citetads{2005MNRAS.358..105J} and the present work, \citetads{2014MNRAS.439.3094B} measured a slow rotation velocity of only $v \sin i = 2$\,km s$^{-2}$ for \object{Proxima}.

The ionized material of the interior of VLMS is a very good electrical conductor, and the interaction of the convective motions with the magnetic field creates Lorentz forces that reduce the velocity of the convection. As a consequence, the efficiency of the energy transportation is reduced and the stellar radius is increased to compensate for the lower effective temperature with an increased radiating surface.
The modeling of the solar-mass binary \object{EF Aqr} by \citetads{2012ApJ...761...30F} showed that the suppression of convection by the magnetic field can effectively alter the structure of solar-type stars.
However, recent modeling of fully convective stars by \citetads{2014ApJ...789...53F} did not demonstrate that the magnetic fields are at the origin of inflated radii for these low-mass objects, unless the internal magnetic fields are extremely strong.
The inflated radii of GJ65 A and B compared to Proxima, while all major physical parameters except rotation rate are identical, is an indication that the required very strong internal magnetic fields may indeed exist in fast-rotating VLMS.

\subsection{Star spots and surface magnetic field\label{starspots}}
Surface magnetic fields are known to locally reduce the efficiency of convection, creating cool spots on the photosphere. If they reach a significant extension, this lowers the average effective temperature of the star and tends to increase its radius.
Following this idea, \citetads{2007ApJ...660..732L} and \citetads{2012ApJ...757...42F} showed a correlation between the X-ray luminosity of the stars and their excess radius compared to model predictions. 
Proxima is known to be very active in the X-ray domain \citepads{2011A&A...534A.133F}, with a quiescent luminosity of $L_X \approx 0.4-1.6 \times 10^{27}$\,erg\,s$^{-1}$ (\citeads{1990A&A...232..387H}, comparable to the solar $L_X$) and flares reaching $L_X = 4 \times 10^{28}$\,erg\,s$^{-1}$ \citepads{2004A&A...416..713G}. Remarkably, it is a member of the UV\,Ceti (=GJ65\,B) class of flaring stars.
This high level of activity may in principle indicate that the surface magnetic field diminishes the efficiency of the surface convection, creating star spots.

However, the quiescent X-ray luminosity of Proxima is comparable to that of GJ65 \citepads{1986MNRAS.219..225A}.
Chandra observations of GJ65 were reported by \citetads{2003ApJ...589..983A}, showing that component B has a stronger magnetospheric activity than component A. This is consistent with the more intense emission lines we observe in the UVES spectrum of B compared to A (Fig.~\ref{uves_emission}).
It therefore appears unlikely that the difference in radius between GJ65\,AB and Proxima is caused by a very different star spot coverage.
Moreover, \citetads{2015ApJ...812...42B} showed using Doppler imaging that M dwarf star spots are typically 200-300\,K cooler than the stellar photospheres and cover only a few percent of their surface.
This moderate coverage implies that star spots probably play a limited role in inflating the radius compared to the (putative) strong internal magnetic fields generated by the dynamo effect.

A survey of the relation between the rotation period and activity in 114 M dwarfs has recently been published by \citetads{2014IAUS..302..176W}. Using the H$\alpha$ emission as a proxy of the activity, the
authors showed a strong correlation of increasing activity with decreasing rotation period.
\citetads{2003ApJ...583..451M} also determined a saturation-type rotation-activity relation, with the saturation level reached around $v \sin i = 10$\,km\,s$^{-1}$ in the M5.5-M8.5 dwarfs.
Spectropolarimetry has allowed considerable progress in the mapping of the magnetic field of stars. Using this technique, \citetads{2010MNRAS.407.2269M} also showed an increase of the large-scale magnetic field strength with decreasing rotation period (their Fig. 15).
These indications suggest that small-scale magnetic field loops at the stellar surface and the associated star spots are enhanced by fast rotation.
But they probably have a limited global effect on the fundamental properties of the star, and in particular on its radius, spin-down, and mass-loss rates \citepads{2014MNRAS.439.2122L}.

\section{Conclusion}
We presented the first interferometric measurements of the angular diameters of the two components of the nearby red dwarf binary GJ65. We also obtained new high-accuracy adaptive optics astrometry and infrared photometry, as well as separate high-resolution spectra of the two stars.
The latter allowed us to derive the differential radial velocity of GJ65 A and B, estimate their rotational velocity $v \sin i,$ and their metallicity with respect to Proxima taken as fiducial.
Based on our new observations, we present refined orbital elements, an accurate value of the total and individual masses of the two components, and of their linear radii.
The positions of GJ65 A \& B in the mass-radius diagram confirms that their radii are underestimated by the current stellar structure models by approximately $13 \pm 4\%$.
Following \citetads{2007A&A...472L..17C}, we propose that the enlargement of their radii is caused by the inhibition of convection by their magnetic fields, generated through dynamo effects by their fast rotation.
This radius inflation is not observed for Proxima, which has almost identical fundamental parameters (in particular the mass) and very similar X-ray activity, but exhibits a slow rotational velocity.

Encouragingly, \citetads{2012ApJ...757...42F} and \citetads{2013ApJ...776...87S} showed that the current generation of VLMS models better agree with observational mass and radius determinations than in the past.
Further progress might be achieved by an improved modeling of the internal magnetic field in fast-rotating fully convective stars. The complexity of the corresponding simulations represents a very difficult challenge, however.
The availability of three VLMS with mostly identical physical properties (GJ65\,AB and Proxima) and differing only in their rotational velocity will be potentially of extremely high value to test their predictions.

\begin{acknowledgements}
We thank J. S. Jenkins for his careful reading of the manuscript, which lead to significant improvements.
We acknowledge financial support from the ``Programme National de Physique Stellaire'' (PNPS) of CNRS/INSU, France.
This research received the support of PHASE, the partnership between ONERA, Observatoire de Paris, CNRS and University Denis Diderot Paris 7.
This research made use of Astropy\footnote{Available at \url{http://www.astropy.org/}} \citepads{2013A&A...558A..33A}, the SIMBAD and VIZIER databases (CDS, Strasbourg, France), NASA's Astrophysics Data System and the Washington Double Star Catalog maintained at the U.S. Naval Observatory.
\end{acknowledgements}

\bibliographystyle{aa} 
\bibliography{biblioUVCet}

\end{document}